\documentclass[11pt,psfig,epsf,bbox]{article}

\begin{document}

\newcommand{\vAi}{{\cal A}_{i_1\cdots i_n}} \newcommand{\vAim}{{\cal
A}_{i_1\cdots i_{n-1}}} \newcommand{\vAbi}{\bar{\cal A}^{i_1\cdots i_n}}
\newcommand{\vAbim}{\bar{\cal A}^{i_1\cdots i_{n-1}}}
\newcommand{\htS}{\hat{S}} \newcommand{\htR}{\hat{R}}
\newcommand{\htB}{\hat{B}} \newcommand{\htD}{\hat{D}}
\newcommand{\htV}{\hat{V}} \newcommand{\cT}{{\cal T}} \newcommand{\cM}{{\cal
M}} \newcommand{\cMs}{{\cal M}^*}
 \newcommand{\vk}{{\bf k}}
\newcommand{\vK}{{\vec K}} \newcommand{\vb}{{\vec b}} \newcommand{{\vp}}{{\vec
p}} \newcommand{{\vq}}{{\vec q}} \newcommand{\vQ}{{\vec Q}}
\newcommand{\vx}{{\vec x}}
\newcommand{\tr}{{{\rm Tr}}} 
\newcommand{\beq}{\begin{equation}}
\newcommand{\eeq}[1]{\label{#1} \end{equation}} 
\newcommand{\half}{{\textstyle
\frac{1}{2}}} \newcommand{\gton}{\stackrel{>}{\sim}}
\newcommand{\lton}{\mathrel{\lower.9ex \hbox{$\stackrel{\displaystyle
<}{\sim}$}}} \newcommand{\ee}{\end{equation}}
\newcommand{\ben}{\begin{enumerate}} \newcommand{\een}{\end{enumerate}}
\newcommand{\bit}{\begin{itemize}} \newcommand{\eit}{\end{itemize}}
\newcommand{\bc}{\begin{center}} \newcommand{\ec}{\end{center}}
\newcommand{\bea}{\begin{eqnarray}} \newcommand{\eea}{\end{eqnarray}}
\newcommand{\beqar}{\begin{eqnarray}} \newcommand{\eeqar}[1]{\label{#1}
\end{eqnarray}} \newcommand{\bra}[1]{\langle {#1}|}
\newcommand{\ket}[1]{|{#1}\rangle}
\newcommand{\norm}[2]{\langle{#1}|{#2}\rangle}
\newcommand{\brac}[3]{\langle{#1}|{#2}|{#3}\rangle} \newcommand{\hilb}{{\cal
H}} \newcommand{\pleft}{\stackrel{\leftarrow}{\partial}}
\newcommand{\pright}{\stackrel{\rightarrow}{\partial}}

\begin{center}
{\Large {\bf{The Ter-Mikayelian Effect on QCD Radiative Energy Loss}}}

\vspace{1cm}

{ Magdalena Djordjevic and Miklos Gyulassy }

\vspace{.8cm}

{\em { Dept. Physics, Columbia University, 538 W 120-th Street,\\ New York,
       NY 10027, USA }} 

\vspace{.5cm}

{July 16, 2003}
\end{center}

\vspace{.5cm}

\begin{abstract}
The color dielectric modification of the gluon dispersion relation
in a dense QCD medium suppresses both the soft and collinear gluon
radiation associated with jet production. We compute
the longitudinal and transverse plasmon contributions to the 
zeroth order in opacity radiative energy loss in the 1-loop HTL approximation.
 This is QCD analog of the 
Ter-Mikayelian effect in QED and leads to $\sim 30\%$ reduction of the energy loss 
of high transverse momentum charm quarks produced 
in a QCD plasma with a characteristic  Debye mass $\mu\sim 0.5$ GeV.
\end{abstract}

\section{Introduction}

 Jet tomography in ultra-relativistic nuclear collisions
can be used to map out the density of the produced QCD
plasma from the suppression pattern of high transverse momentum
hadrons~\cite{TOMO}-\cite{Wang}. The quenching of 
jets~\cite{Gyulassy:1990bh,Gyulassy:1991xb} is mainly  due to the medium 
induced radiative energy loss of high energy partons
propagating through ultra-dense QCD matter. However, even if 
final state multiple elastic and inelastic interactions
are neglected, gluon radiation associated with hard QCD
processes  already softens considerably the lowest order jet spectra.  
In elementary particle collisions,
this {\em associated}  radiation can be 
taken into account through the $Q^2$ (DGLAP) evolution of the hadronic
fragmentation functions. In a plasma, even the associated radiation is modified by the
dielectric properties of that medium as first pointed out by Ter-Mikayelian~\cite{TM1, TM2}.
The non-abelian QCD analog of the Ter-Mikayelian plasmon
effect is the subject of this paper. A summary of our main results has been
published in paper I\cite{DG-PLBCharm}. In this paper, the details of 
Ter-Mikayelian calculations are  presented.

This work is motivated by the surprising observation of PHENIX~\cite{Adcox:2002cg} that ``prompt'' single electron spectrum from
open charm production in $Au + Au$ collisions at $\sqrt{s}=130$ AGeV
shows no sign of heavy quark energy loss~\cite{Batsouli_Gyulassy}.  In
contrast, a dramatic suppression (by a factor $\sim 5$) was observed at RHIC
for pions with $p_T>3$ GeV originating from the fragmentation of light quark and gluons
\cite{Adler:2002xw1}-\cite{brahms_03}. See a recent reviews of
light quark and gluon tomography in A+A in~\cite{Gyulassy:2003mc}-\cite{Kovner:2003zj}.
The suppression of light hadrons is consistent with the expected large
radiative induced energy loss of light quark and gluon jets in an
ultra-dense plasma of density approximately 100 times higher than in
ground state nuclei~\cite{Vitev:2002pf}.

The spectrum of induced  radiation depends on the optical thickness or opacity
$\chi=L/\lambda$ of the plasma, and has been computed to arbitrary  order in
$\chi^n$ for massless partons in GLV~\cite{GLV,Gyulassy:2003mc}. 
It was expected~\cite{Shuryak:1996gc}-\cite{Lin:1998bd} 
that a similar quenching pattern 
should also be observed for heavy
quark ($c$ or $b$) jet fragmentation. 
However, in~\cite{Dokshitzer:2001zm} it was
pointed out that the a large quark mass would lead to a ``dead cone'' effect,
reducing induced radiation inside the cone $\theta<M/E$ 
and that this should reduce radiative energy loss of
heavy quarks as compared to light partons.  Numerical estimates
indicated that the quenching of heavy charm quarks may be only 
about a half that of light quarks.  PHENIX data,
however, suggest that the charm quark energy loss could be even smaller
than that.
In I~\cite{DG-PLBCharm} we showed that the apparent null effect
observed for heavy quark energy loss via single electrons may in part be due
to a further reduction of the leading order $O(\chi^0)$ {\em associated}.
In this sense, there are 
two opposing medium effects: (1) at $O(\chi^0)$ the Ter-Mikayelian (plasmon) effect 
that arises at 1-HTL loop level the reduces the associated energy loss 
and (2) at $O(\chi^{n\ge 1})$ the 
 induced radiative energy loss that increases the total radiative energy loss. 
Our method of calculating the
Ter-Mikayelian effect in QCD is discussed in detail below.

In~\cite{Dokshitzer:2001zm} the absence of radiation below a plasma
frequency cutoff was estimated to reduce the
induced energy loss by only $\sim 10\% $.  However, the Ter-Mikayelian effect on the zeroth order  in opacity $(L/\lambda)^0$ {\em associated} radiation 
was not considered up to now.  The first estimates of the influence of a
plasma frequency cutoff in QCD plasmas were reported 
in~\cite{Kampfer_Pavlenko, Zakharov} using a constant plasmon mass
$\omega_{0}$~\cite{Biro_Levai_Muller}-\cite{Levai_Heinz}. The $k$
dependence of the gluon self energies and the magnitude of
longitudinal radiation were not investigated in that work. We
extend those results by taking both longitudinal as well as transverse
modes consistently into account via the frequency and wavenumber
dependent hard thermal (1-loop HTL) self energy
$\Pi^{\mu\nu}(\omega, \vec{\mathbf{k}})$~\cite{Kalashnikov:cy}-\cite{Gyulassy_Selikhov}.

As noted in I~\cite{DG-PLBCharm}, the dielectric properties of an isotropic  plasma
lead to a transverse gluon self energy $\Pi_T(\omega,\vec{\mathbf{k}} )$ 
with $\Pi_T(\omega_{pl}(0),0)=\omega_{pl}^2(0)\approx \mu^{2}/3$, 
where $\mu\approx gT$ is the Debye screening mass of a plasma at temperature
$T$ in the deconfined phase. In addition, 
long wavelength collective longitudinal gluon modes arise with $\Pi_L(\omega_{pl}(0),0)=
\omega_{pl}^2(0)$. This dynamical gluon mass, $\surd(\omega_{pl}^2(k)-k^2)$,
suppresses not only the radiation of sub-plasmon $\omega<\omega_{pl}(|\vec{\mathbf{k}}|)$ 
gluons but also shields against the collinear $k_\perp\rightarrow 0$ singularities. 

In this paper we study under what conditions the assumption of using a $k$ independent
effective plasmon mass and neglecting longitudinal modes may be adequate.
We show below  that this simplifying assumption is surprisingly accurate ($\sim 10 \%$)
if the asymptotic mass~\cite{Rebhan}, $\omega_\infty=m_E/\sqrt{2}$ rather than the $k=0$ $\omega_{pl}=\sqrt{2/3} \omega_{\infty}$ is employed. The accuracy 
of the approximation also improves dramatically as the mass of the heavy quark
increases. 

\section{Jet Production in the Vacuum}

In order to introduce the method that we use to  compute  the
QCD analog of the Ter-Mikayelian~\cite{TM1}-\cite{TM2} effect on the zeroth order in opacity radiation, we consider first the well known case of radiation in the vacuum.

\begin{center}
\vspace*{1.0cm}
\includegraphics{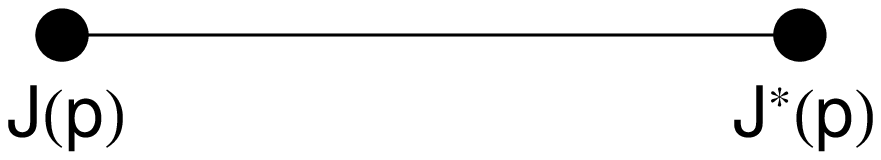}  
\vskip -10pt
{\small {FIG.~1} Illustration of a bare jet vacuum to vacuum amplitude.}
\end{center}

The vacuum to vacuum amplitude, illustrated in Fig.1, is given by 
\beq
-iM_0= \frac{d_J}{2} \int \frac{d^4 p}{(2\pi)^4} |J(p)|^{2} \Delta_M(p)
\eeq{m0}
where $\Delta_M(p)=(p^2+M^2+i\epsilon)^{-1}$ is the jet propagator
for a spinless parton of mass $M$ in the $d_J$ dimensional representation
of $SU(N_c)$. We will ignore spin effects throughout
since they are irrelevant in the soft radiation limit.
The effective jet source current, $J$, creates an invariant distribution of 
jets as given by 
\beq 
2 {\rm Im} M_0= \int d^3 N_J= 
\int \frac{d^3 \vec {\bf p}}{E_p} \; E_p\frac{d^3 N_J}{d^3 \vec {\bf p}},
\eeq{dnj}
where $E_p d^3N_J/d^3 \vec {\bf p}=d_J |J(E_p, \vec {\bf p})|^2/(2(2\pi)^3)$ with $E_p^2=M^2+ \vec {\bf p}^2$.

\begin{center}
\vspace*{5.5cm}
\includegraphics{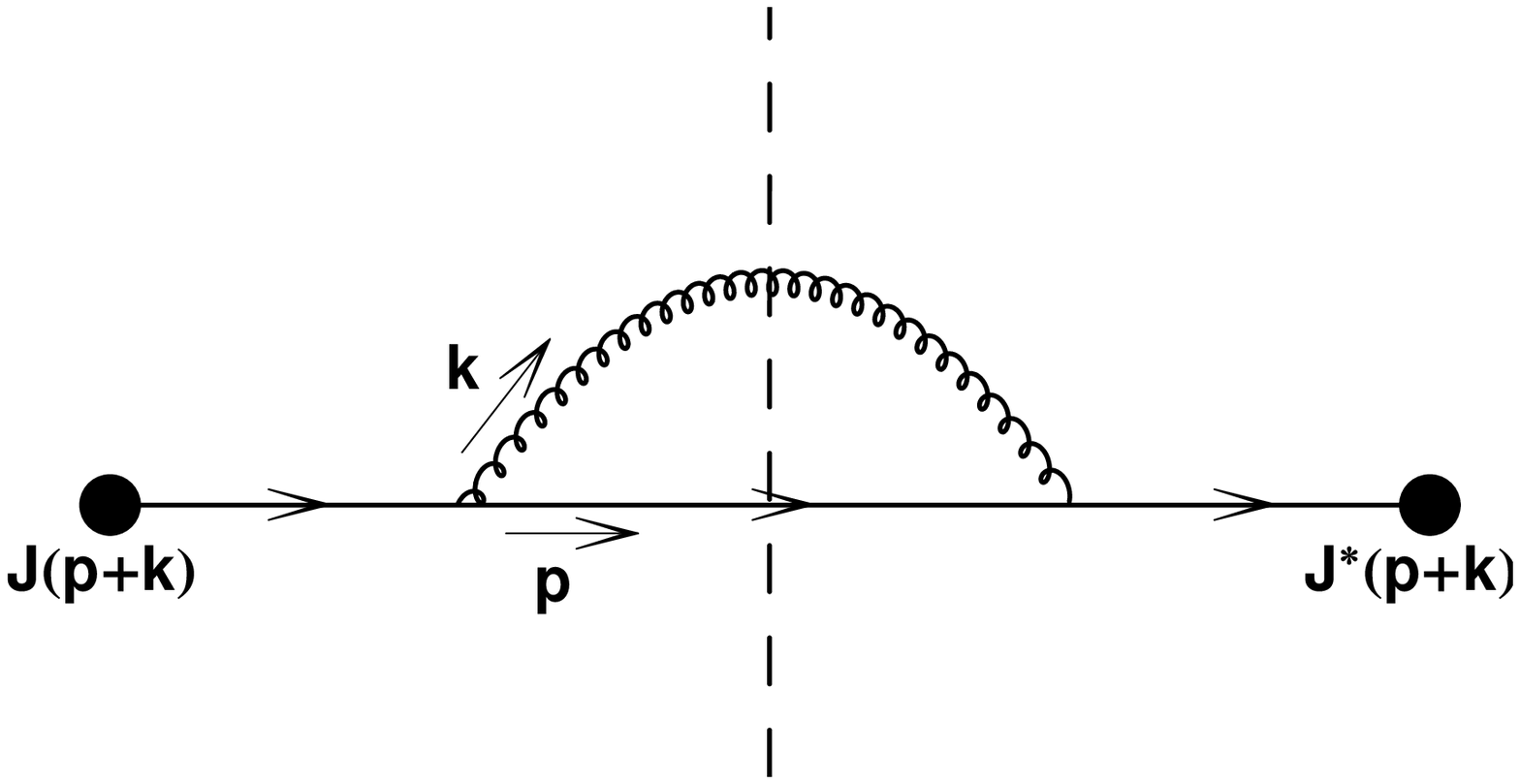}  
\vskip 33pt
{\small {FIG.~2} The one gluon {\em associated} radiation amplitude in the vacuum.}
\end{center}

The one gluon radiative correction amplitude, $iM_1$, to the jet spectrum
in the vacuum is illustrated on Fig.2 
\beq
-iM_{1} =- \frac{g_{s}^{2} C_J d_J}{2} \int \frac{d^{4}p}{(2\pi )^{4}}
\frac{d^{4}k}{(2\pi )^{4}}|J(p+k)|^2 \Delta_M(p+k)^2 \Delta_M(p)
(2p+k)^{\mu }D^{(0)}_{\mu \nu }(k)(2p+k)^{\nu }.
\eeq{m1}
The free gluon propagator in the axial ($u_\mu A^\mu=0$) gauge is 
\beq
D^{(0)}_{\mu\nu}(k)= -\left( g^{\mu\nu}- \frac{u^\mu k^\nu +
 k^\mu u^\nu}{(uk)}+ u^2 \frac{k^{\mu} k^\nu}{(uk)^2} \right)\Delta_{m}(k)
\; \; , \eeq{d0}

where $m$ is an infrared mass in the massive gluon scheme introduced by need 
to study the $m=0$ limit. In the next section, the dynamic polarization 
tensor will replace $m$.

The imaginary part of this amplitude contains real and virtual radiation 
corrections to $iM_0$. The inelastic one gluon radiation contribution
is obtained by the Cutkosky rule~\cite{Peskin}
\beq
\Delta_M(p)\Delta_{m}(k) \rightarrow (-2\pi i)^2 \delta(k^2-m^2)
\delta(p^2-M^2)
\; \; .
\eeq{cut1}
This gives a contribution the jet plus gluon production rate
\beqar
2 {\rm Im}M_1|_{rad}&=& \int \frac{d^{4}p}{(2\pi )^{3}}\delta(p^2-M^2)d_J
\int \frac{d^{4}k}{(2\pi )^{3}}\delta(k^2-m^2) |J(p+k)|^2 
\nonumber \\
&\; &  
 \hspace{0in} \times \frac{C_J g_s^2}{(Q^{2}-M^{2})^{2}} 
\left \{(Q^{2}-M^{2})\frac{4(pu)}{(uk)}- 
(Q^{2}-M^{2})^{2} \frac{u^2}{(uk)^2}-4M^2+m^2\right \},
\eeqar{imm1}
where
\beq
Q^{2}=(p+k)^{2}=2(pk)+k^{2}+M^{2}.
\eeq{Q^2}

In general, the result depends on the gauge parameter $u^\mu$
because the external color source $J$ breaks gauge invariance. However, in 
the soft gluon limit, $k_\perp\ll k^+\ll p^+$,
where $k_\perp$ is measured relative to the $\vec {\bf p}$ axis of the jet,
one can extract the familiar  DGLAP soft radiation spectrum
with any $u^\mu$ for which $(uk)/(up)\approx k^+/p^+ =x\ll 1$.
For typical light cone kinematics of interest, the momenta are expressed as
\beqar
k^\mu&=&[x E^+, (k_\perp^2+m^2)/xE^+, \vk_\perp]\nonumber \\
p^\mu&=&[(1-x)E^+,(M^2+k_\perp^2)/(1-x)E^+,-\vk_\perp]
\nonumber \\
2pk&=& \frac{k_\perp^2+(1-x)^2 m^2+x^2 M^2}{x(1-x)}
\nonumber \\
Q^{2}&=&pk+m^2 + M^{2} =  \frac{k_\perp^2+(1-x) m^2+x M^2}{x(1-x)}
\; \; . 
\eeqar{kinem}
Note that the vertex factor $\{ ... \}$ in Eq. (\ref{imm1})
reduces to $(2pk+m^2)/x$ in the $x\rightarrow 0$ limit
in both the $A^+=0$ light cone and temporal $A^0=0$ gauges.

We assume that $J$ is slowly varying $J(p+k)\approx J(p)$ for soft radiation, 
and therefore, in soft radiation approximation the spectrum can be extracted  
as
\beqar
2 {\rm Im}M_1|_{rad}&\approx& \int d^{3}N_J \int d^3N_g^{(0)}
\; \; ,\eeqar{imm1a}
leading to the  finite mass generalization of the small $x$ invariant DGLAP
radiation spectrum 

\beqar
\omega \frac{dN_g^{(0)}}{d^3\vk} \approx x\frac{dN_g^{(0)}}{dx d^2\vk_\perp}&
\approx& \frac{C_{J} \alpha_{s}}{x \pi^{2} (Q^{2}-M^{2})} = 
\frac{C_J \alpha_s }{\pi^2} \frac{1}{k_\perp^2+ (1-x) m^2+x^2 M^2} 
\; \; . 
\eeqar{imm2}

\section{Ter-Mikayelian effect} 

In this section we want to compute (zeroth order in opacity) associated radiative 
quark energy loss. In soft gluon limit, the result should not depend on the 
choice of gauge, as long as $(uk)/(up) \approx x \ll 1$ is satisfied. We 
simplify 
our calculations by choosing the temporal axial gauge. However, we have 
explicitly checked that, in soft gluon limit, the same result is 
obtained using the light cone gauge. 

The radiative heavy quark energy loss in hot dense medium involves both
transverse and longitudinal gluon radiation. In temporal axial gauge gluon propagator has the following form (see appendix A):
\beq
D_{\mu \nu }=-\frac{P_{\mu \nu }}{\omega ^{2}\epsilon _{T}-
\overrightarrow{\mathbf{k}}^{2}}-\frac{Q_{\mu \nu }}{\omega ^{2}\epsilon _{L}},
\eeq{dmnMed}
where transverse $(P_{\mu \nu })$ and longitudinal $(Q_{\mu \nu })$
projectors are given in terms of $\overline{g}_{\mu \nu }=g_{\mu \nu }-\frac{k_{\mu }k_{\nu }}{k^{2}}$ and $\overline{u}_{\mu }=\overline{g}_{\mu \nu }u_{\nu }$, by
\beq
P_{\mu \nu }=\overline{g}_{\mu \nu }-\frac{\overline{u}_{\mu }
\overline{u}_{\nu }}{\overline{u}^{2}},
\eeq{Pmunu}
\beq
Q_{\mu \nu }=(g_{\mu \nu }-\frac{u_{\mu }k_{\nu }+k_{\mu }u_{\nu }}
{(u\cdot k)}+\frac{u^{2}k_{\mu }k_{\nu }}{(u\cdot k)^{2}}-P_{\mu \nu })
\frac{(u \cdot k)^{2}}{k^{2} u^{2}}
\eeq{}
and 
\beq
\epsilon _{T}=1-\frac{\mu^{2}}{2\vec{\mathbf{k}}^{2}}
(1-\frac{\omega ^{2}-\vec{\mathbf{k}}^{2}}{2\omega 
|\vec{\mathbf{k}}|} \log (\frac{\omega +
|\vec{\mathbf{k}}|}{\omega -|\vec{\mathbf{k}}|})),
\eeq{e_T}
\beq
\epsilon _{L}=1+\frac{\mu^{2}}{\vec{\mathbf{k}}^{2}}
(1-\frac{\omega }{2|\vec{\mathbf{k}}|}\log 
(\frac{\omega +|\vec{\mathbf{k}}|}
{\omega -|\vec{\mathbf{k}}|}))
\eeq{e_L}
are transverse and longitudinal dielectric functions respectively~\cite{Gyulassy_Selikhov}.

\bigskip 

In order to compute the associated radiative energy loss when the hard process is embedded in a dielectric medium, we need to compute
the squared amplitude of Feynman diagram diagram $|M_{rad}|$, which 
represents the source $J$ that produces an off-shell jet with
momentum $p^{\prime }$, which subsequently radiates an  gluon with 
obeying the dispersion relation, $\omega(k)$, of the medium with momentum $k$. 
The jet emerges with momentum $p$ and mass $M$. Since our focus is on heavy quarks,
we neglect the thermal shifts of the heavy quark mass here. To 
calculate this rate  we use the optical theorem via
\beq
\int |M_{rad}|^{2}\frac{d^{3} \vec {\bf p}}{(2\pi )^{3}2E}
\frac{d^{3} \vec {\bf k}}{(2\pi)^{4}2\omega }=2 {\rm Im}M_{TM}= \int d^{3}N_J 
\int d^3N_g^{TM} ,
\eeq{}
where, in axial gauge, $Im M_{TM}$,  is the imaginary part one Hard Thermal Loop 
amplitude of the the diagram cut as shown in Fog.3. 
\begin{center}
{Im$M_{TM}$}
\end{center}
\begin{center}
\vspace*{5.4cm}
\includegraphics{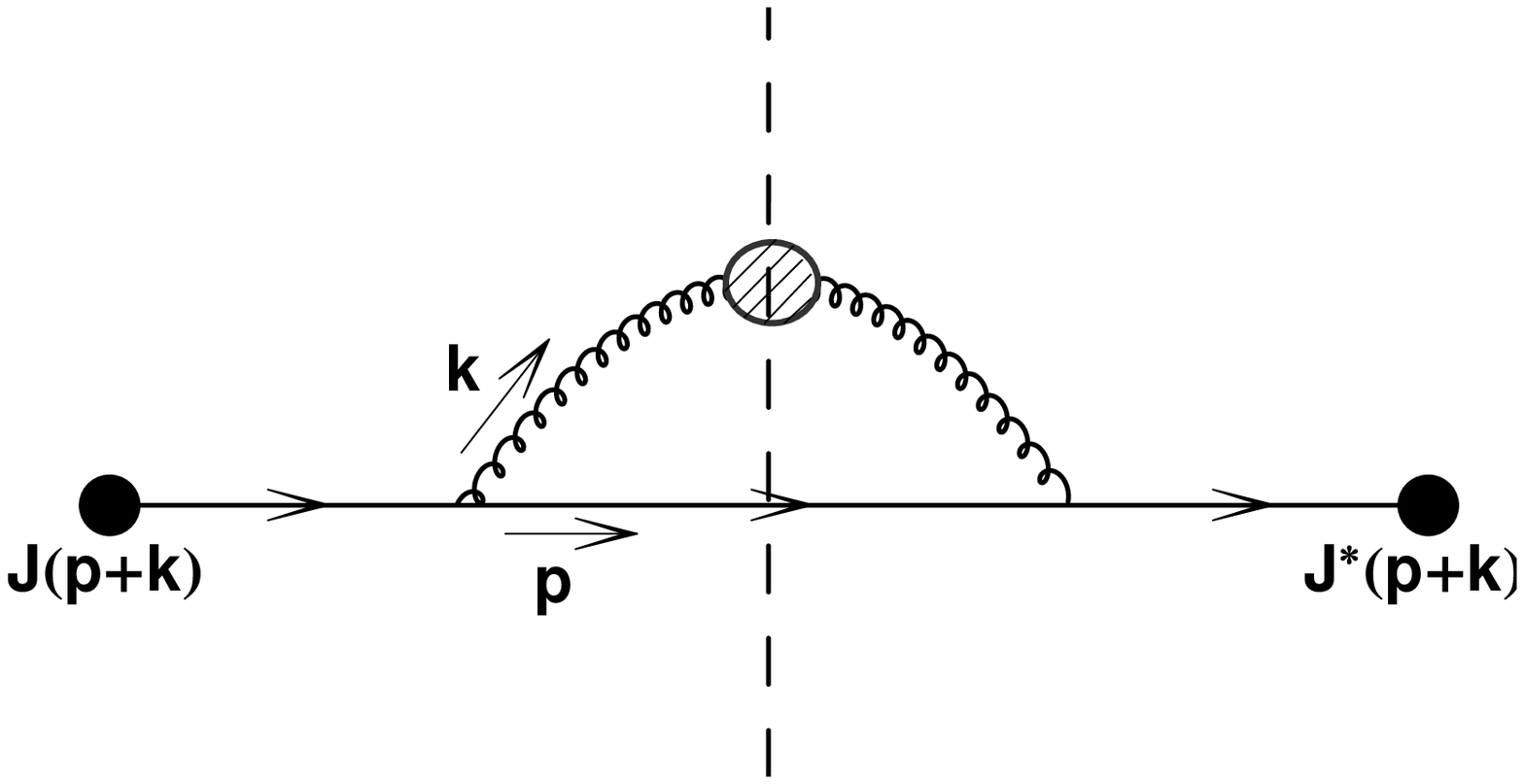}  
\vskip 25pt
\begin{minipage}[t]{15.0cm}
{\small {FIG.~3} shows the cut amplitude that contributes to 
the associated radiative energy loss in the medium . The dashed line 
shows which propagators are to be put on-shell with respect to the dispersion
in the medium. The blob denotes the resumed 
HTL self energy using  Eq. (\ref{dmnMed}, \ref{e_T}, \ref{e_L}). }
\end{minipage}
\end{center}

We assume, as in ~\cite{GLV}, that $J$ varies slowly with $p$, i.e. 
that $ J(p+k)\approx J(p)$ and neglect high $x$  spin effects. 
amplitude in Fig.3 before cutting is
\beq
-iM =-\frac{C_{R}D_{R}}{2}\int 
\frac{d^{4}p}{(2\pi )^{4}} \frac{d^{4}k}{(2\pi )^{4}}|J(p)|^{2}
g_{s}^{2} \frac{1}{(Q^{2}-M^{2})^2}\frac{1}{p^{2}-M^{2}+i\epsilon}
(2p+k)^{\mu }D_{\mu \nu }(2p+k)^{\nu }.
\eeq{}
The contribution to medium on-shell radiation is 
\beq
2 {\rm Im}M_{TM}=i C_{R}D_{R} \int \frac{d^{3}p}{(2\pi )^{3}2E} |J(p)|^{2}
 \frac{d^{4}k}{(2\pi )^{4}} g_{s}^{2} 
\frac{1}{(Q^{2}-M^{2})^{2}} (2p+k)^{\mu }D_{\mu \nu }(2p+k)^{\nu}
\eeq{}
where we have used~\cite{Peskin} $1/(p^{2}-M^{2}+i\epsilon) \rightarrow (-2\pi i)\delta(p^{2}-M^{2})$ for the cut quark propagator in Fig.3 (neglecting its small thermal shift). 
As shown in Appendix B,, the contour integration over $\omega$ 
for on-shell quark and glue is equivalent to replacing $D_{\mu \nu }$ by:
\beqar
D_{\mu \nu } \rightarrow -P_{\mu \nu }(-2\pi i)\delta (\omega ^{2}
\epsilon _{T}-\vec{\mathbf{k}}^{2}) - \frac{Q_{\mu \nu }}{\omega ^{2}}
(-2\pi i)\delta (\epsilon _{L}), 
\eeqar{Dmunu} 
where we only keep positive energy contribution.
A small $O(m_g/E)$ quasi elastic contribution is neglected in this approximation as discussed in App. B. 
We obtain in this way 
\beqar
\int d^{3}N_J \int d^3N_g^{TM}&=&\int \frac{d^{3} \vec {\bf p}}
{(2\pi )^{3}2E} D_{R} |J(p)|^{2} \int C_{R} g_{s}^{2} \frac{d^{4}k}
{(2\pi )^{3}} \frac{1}{(Q^{2}-M^{2})^{2}} \times \nonumber \\
&\;& \times (2p+k)^{\mu }(-P_{\mu \nu }\delta (\omega
^{2}\epsilon _{T}-\vec{\mathbf{k}}^{2})-
\frac{Q_{\mu \nu }}{\omega ^{2}}\delta (\epsilon _{L}))(2p+k)^{\nu}
\eeqar{}

It is convenient to choose coordinates such 
that $p=(E,|\vec{\mathbf{p}}|,0,0)$, where $E=\sqrt{\vec{\mathbf{p}}^{2}+
M^{2}}$, and $k=(\omega ,|\vec{\mathbf{k}}|\cos \theta ,|\vec{\mathbf{k}}|
\sin \theta ,0)$.
\medskip
With this kinematics we get:
\beqar
&&(2p+k)^{\nu }P_{\nu \rho }(2p+k)^{\rho }= -(2p+k)^{i}(\delta_{ij}- 
\frac{k_{i}k_{j}}{\vec{k}^{2}})(2p+k)^{j}
=-4\vec{\mathbf{p}}^{2}\sin ^{2}\theta 
\nonumber \\
&&(2p+k)^{\nu }Q_{\nu \rho }(2p+k)^{\rho }= -(2p+k)^{i}(\frac{k_{i}k_{j}}
{\vec{k}^{2}})(2p+k)^{j}=-(2|\vec{\mathbf{p}}|\cos \theta +
|\vec{\mathbf{k}}|)^{2} \; . 
\eeqar{PandQ}

The integrated radiation yield than becomes:

\beqar
\int d^3N_g^{TM}&=& \int d\omega d\cos \theta 
\vec{\mathbf{k}}^{2}d|\vec{\mathbf{k}}| \frac{C_{R} \alpha_{S}}{\pi} \frac{1}{(Q^{2}-M^{2})^{2}}\, \times
\nonumber \\
&\times& \{4\vec{\mathbf{p}}^{2}\sin^{2}\theta 
\delta (\omega ^{2}\epsilon _{T}-\vec{\mathbf{k}}^{2})+ 
\frac{(2|\vec{\mathbf{p}}|\cos \theta +
|\vec{\mathbf{k}}|)^{2}}{\omega ^{2}} \delta (\epsilon_{L}) \}
\eeqar{yield}

where $Q^{2}-M^{2}=\omega ^{2}-\vec{\mathbf{k}}^{2}+
2(E \omega-|\vec{\mathbf{p}}||\vec{\mathbf{k}}|\cos \theta)$, as 
in Eq.~(\ref{Q^2}).

\medskip

The Ter-Mikayelian modified associated (0$^{th}$ order in opacity) radiative energy loss 
spectrum is 
defined as  $dI_{g}=\omega d^3N_g^{TM}$. Using this, the transverse and 
longitudinal contribution to the 0$^{th}$ order radiated energy loss per 
wave number is given by the following:
\beqar
\frac{dI_{T}}{d|\vec{\mathbf{k}}|}&=&\frac{C_{F}}{\pi } \;
\frac{4  {\vec{\mathbf{k}}}^{2}
\vec{\mathbf{p}}^{2} \omega _{T}^{2} (\omega _{T}^{2}-\vec{\mathbf{k}}^{2})}
{\omega _{T}^{2}\mu^{2}-(\omega _{T}^{2}-\vec{\mathbf{k}}^{2})^{2}}
\int_{0}^1 d\cos\theta \; \frac{\alpha _{S}(Q^2-M^{2})}{(Q^2-M^2)^2} \sin^2\theta
 \nonumber \\
\frac{dI_{L}}{d|\vec{\mathbf{k}}|}&=&\frac{C_{F}}{\pi }
\frac{4 \vec{ \mathbf{k}}^{2}
\vec{\mathbf{p}}^{2}(\omega _{L}^{2}-\vec{\mathbf{k}}^{2})}
{\mu^{2}-(\omega _{L}^{2}-\vec{\mathbf{k}}^{2})}
\int_{0}^1 d \cos \theta \; \frac{\alpha _{S}(Q^2-M^{2})}{(Q^2-M^2)^2} 
\left(\cos \theta + \frac{\vec{\mathbf{k}}^{2}}{2 \vec{\mathbf{p}}^2}\right)^2
\eeqar{en_losses}
where we keep only the forward, $\theta > 0$, emission to isolate the energy 
loss of the nearside jet. In Eq. (\ref{en_losses}) $\omega_{T}$ and $\omega_{L}$ are positive zeros of $(\omega^{2} \epsilon_{T}-\vec{\mathbf{k}}^{2})$ and $\epsilon_{L}$ respectively.
\medskip
The angular integration can be performed analytically if $\alpha_S$ does 
not run, but it is not particularly useful. We perform the integration using 
no momentum cutoffs and running coupling according to the
\smallskip
"Frozen $\alpha$ model"~\cite{Dokshitzer-FrozenAlpha} 
\beq
\alpha_{S}(Q^{2}-M^{2})={\rm Min} \{ 0.5, \frac{4 \pi}
{\beta_{0} Log(\frac{Q^{2}-M^{2}}{\Lambda_{QCD}^{2}})} \}  \;\;\;,
\eeq{frozen_alpha}
where $\beta_{0}=\frac{28}{3}$ for effective number of flavors 
$n_{f} \approx 2.5$ and $\Lambda_{QCD} \approx 0.2$ GeV. 
\medskip
Figures 4 and 5 show transverse an longitudinal contribution to the charm and bottom radiative energy loss at different Debye masses.

\begin{center}
\vspace*{5.5cm}
\includegraphics{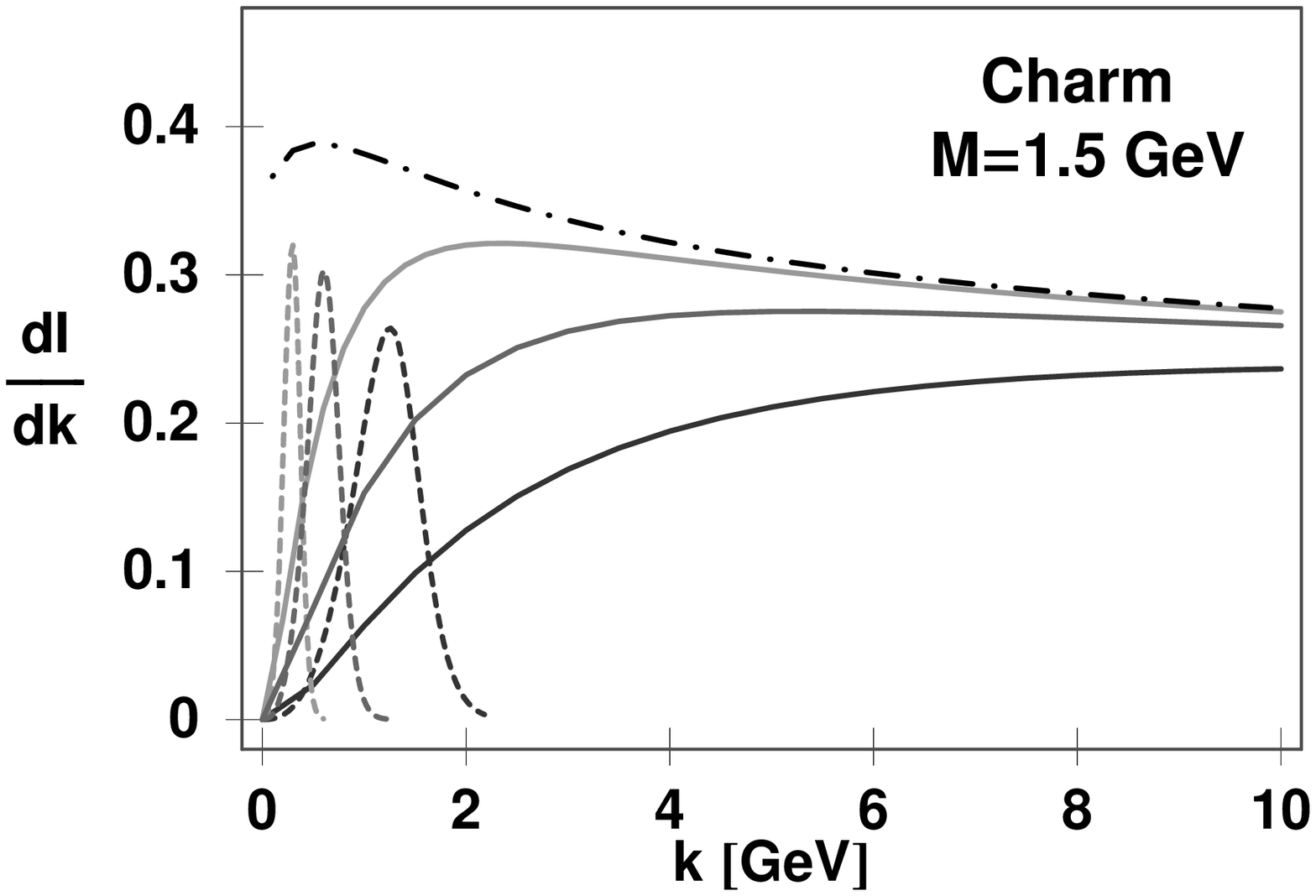}  
\vskip 42pt
\begin{minipage}[t]{15.0cm}
{\small {FIG.~4} The $0^{th}$ order in opacity contribution to Charm quark 
radiated energy loss for $15$ GeV jet is shown as a function of wave number. The dashed-dotted curve shows what would the energy loss be if gluons were treated as massless and transversely polarized. From top to bottom (left to right) solid (dashed) curves show medium modified transverse   (longitudinal) contribution to the energy loss for Debye mass $0.25$ GeV, $0.5$ GeV and $1$ GeV respectively.}
\end{minipage}
\end{center}
\vskip 4truemm 

\begin{center}
\vspace*{5cm}
\includegraphics{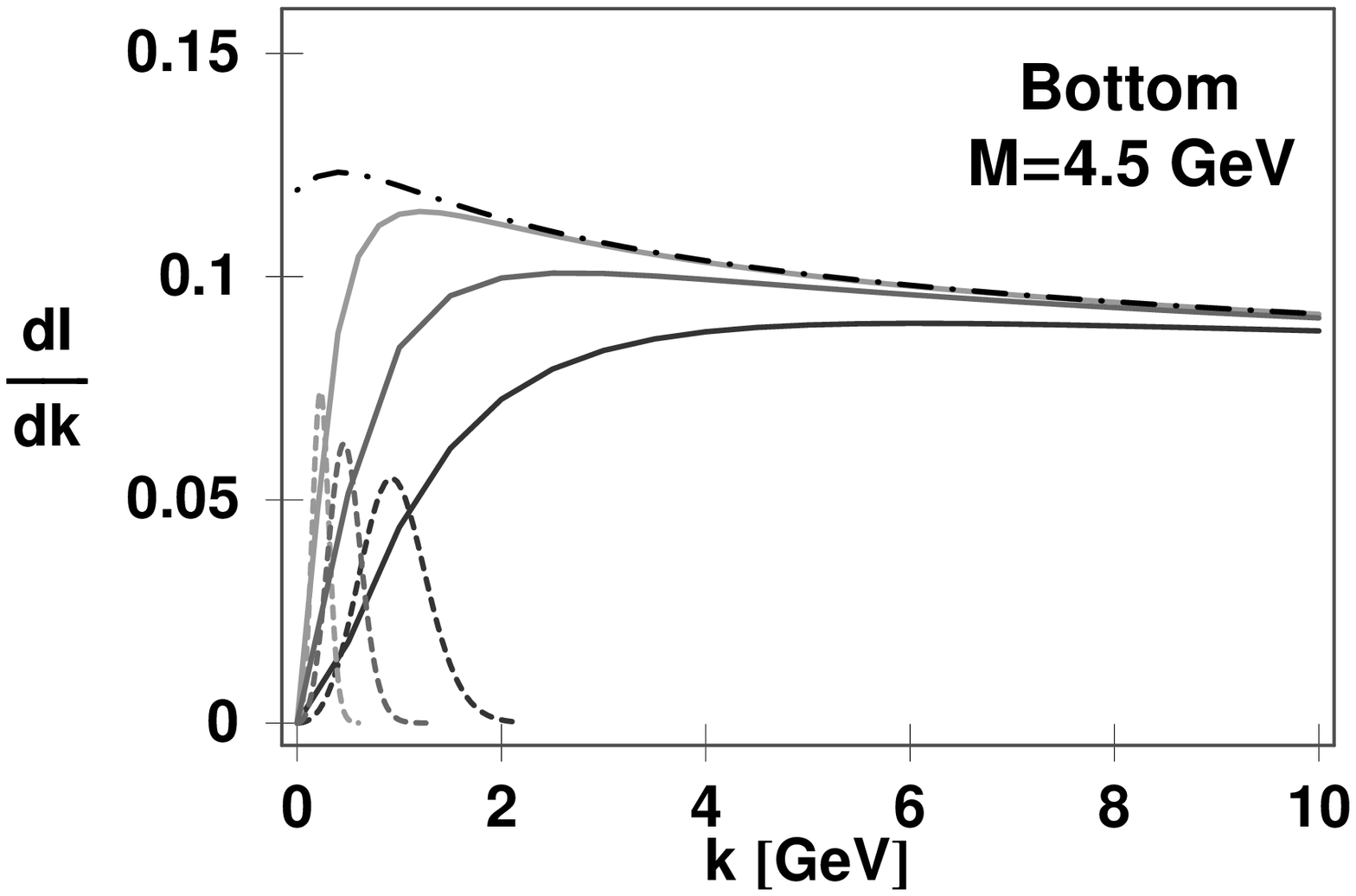}  
\vskip 42pt
\begin{minipage}[t]{15.0cm}
{\small {FIG.~5} The $0^{th}$ order in opacity contribution to Bottom 
quark radiated energy loss for $15$ GeV jet is shown as a function of wave number. The dashed-dotted curve shows what would the energy loss be if gluons were treated as massless and transversely polarized. From top to bottom (left to right) solid (dashed) curves show medium modified transverse   (longitudinal) contribution to the energy loss for Debye mass $0.25$ GeV, $0.5$ GeV and $1$ GeV respectively.}
\end{minipage}
\end{center}
\vskip 4truemm

We see that longitudinal contribution to the energy loss is significant only in the $\omega $ region around Debye mass. As temperature T of the medium increases the transverse contribution to the energy loss decreases, while longitudinal contribution increases. Therefore, we can conclude that for lower Debye masses (lower temperatures) the longitudinal contribution to the energy loss is negligible, but at high enough temperatures it may become comparable to the transverse contribution.

\medskip

Fig.6 shows the integrated $0^{th}$ order fractional energy loss for charm and bottom
quarks in a hot plasma with Debye mass $\mu=0.5$ GeV.

\begin{center}
\vspace*{5.9cm}
\includegraphics{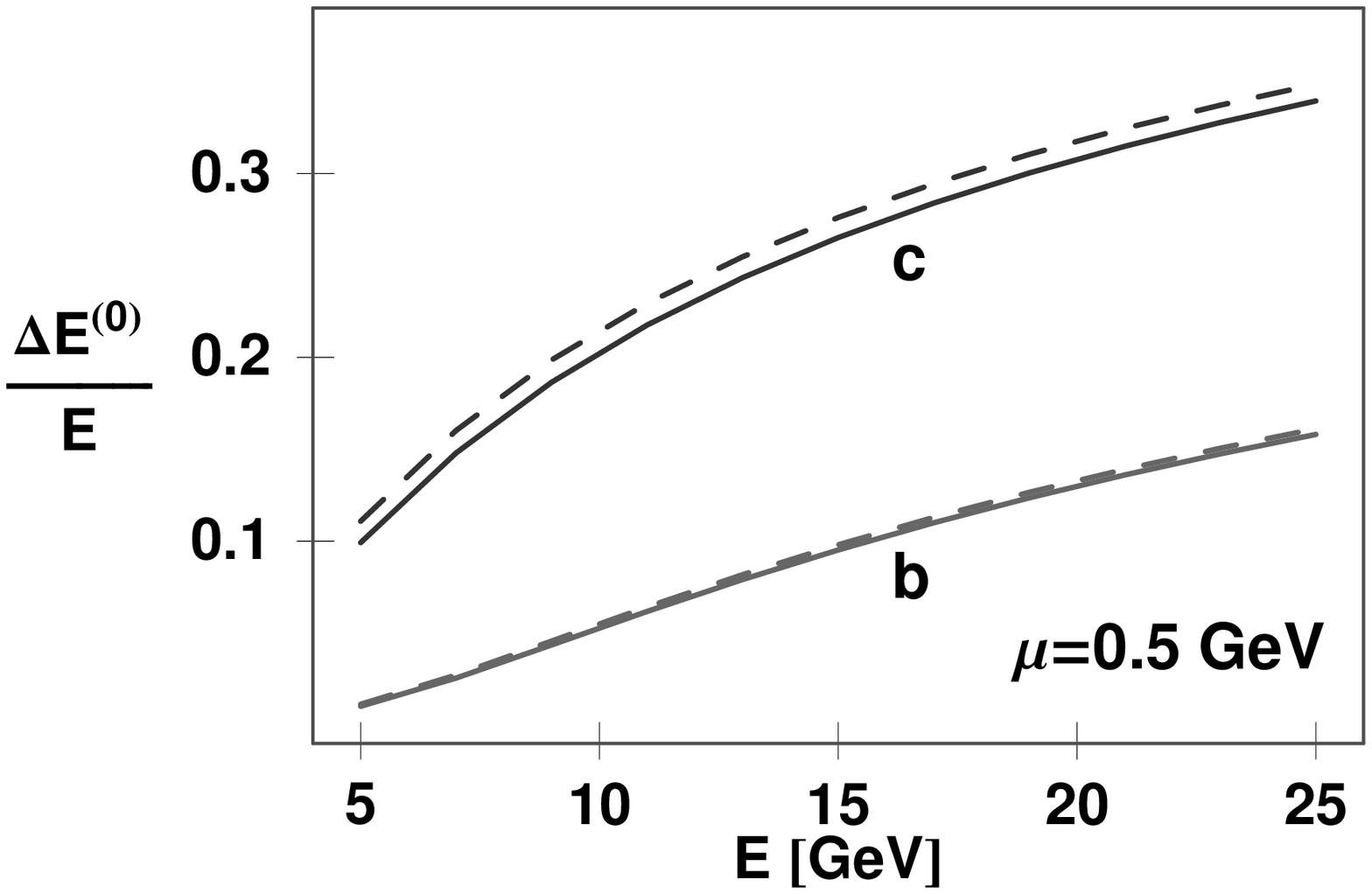}  
\vskip 38pt
\begin{minipage}[t]{15.0cm}
{\small {FIG.~6} The $0^{th}$ order in opacity fractional energy loss for 
charm and bottom quarks in hot medium with Debye mass $\mu=0.5$ GeV, and 
zero momentum cutoff is shown as a function of the charm quark energy. The upper (lower) solid curve shows transverse fractional energy loss for charm 
(bottom) quark. The dashed curves show the negligible additional effect of 
longitudinal plasmons.}
\end{minipage}
\end{center}
\vskip 4truemm 
We see that the longitudinal plasmon contribution is indeed negligible
for both charm and bottom quarks.

In order to study in more detail how the specific 1-loop HTL dispersion relation affects
the results we show in Fig. 7 the dynamical gluon mass as a function of wavenumber for different screening scales.

\begin{center}
\vspace*{6cm}
\includegraphics{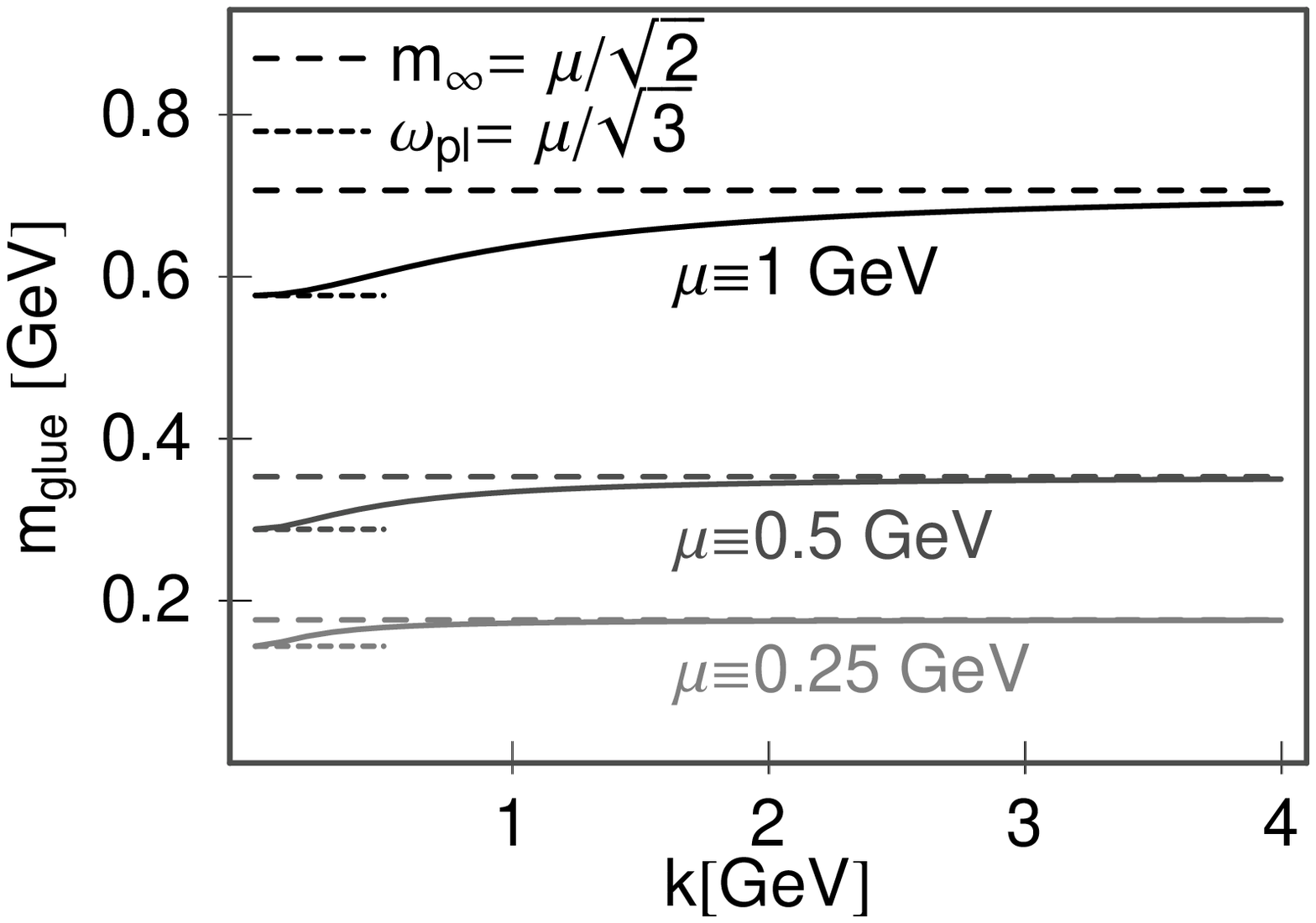}  
\vskip 44pt
\begin{minipage}[t]{15.0cm}
{\small {FIG.~7} One loop transverse plasmon mass $m_{g}(|\vec{\bf{k}}|)\equiv 
\sqrt{\omega ^{2}-\vec{\bf{k}}^{2}}$ is shown as a function of quark momentum 
$|\vec{\bf{k}}|^{2}$. We see that $m_{g}$\ starts with the 
value $\omega_{pl}=\mu/\sqrt{3}$ at low $|\vec{\bf{k}}|$, and that 
as $|\vec{\bf{k}}|$ grows, $m_{g}$ asymptotically approaches the value 
of $\omega _{\infty }=\mu/\sqrt{2}$ in agreement with~\cite{Rebhan}.} 
\end{minipage}
\end{center}
\vskip 4truemm 
The main feature to note is the relatively rapid transition from
the static $\omega_{pl}=\mu/\sqrt{3}$ to an asymptotic $m_{\infty}=\mu/\sqrt{2}$
as emphasized in \cite{Rebhan}. The effect of replacing the dynamical mass by
a single effective value, either $\omega_{pl}, m_{\infty}$, or $\mu$ is shown in Fig.8. 

\begin{center}
\vspace*{6cm}
\includegraphics{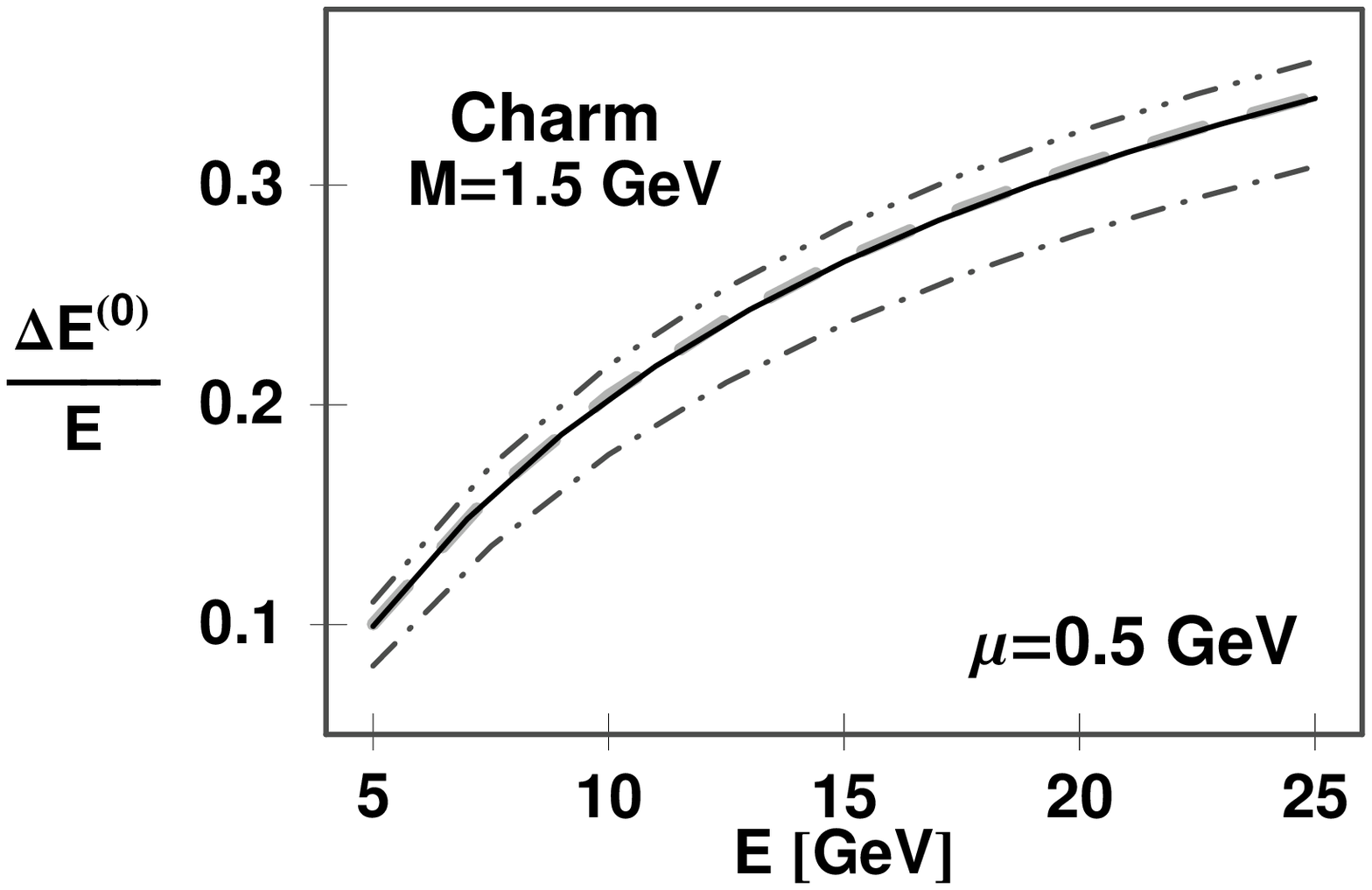}  
\vskip 44pt
\begin{minipage}[t]{15.0cm}
{\small {FIG.~8} Medium modified zeroth order in opacity charm quark 
fractional energy loss is shown as a function of quark energy. Full curve 
shows the transverse energy loss using Eq. (\ref{en_losses}). Dot-dot-dashed, 
dashed and dot-dashed curves show what would be the transverse energy loss if 
we define gluon mass as $\omega_{pl}$, $m_{infty}$ and $\mu$ respectively.} 
\end{minipage}
\end{center}
Figs. 7,8 demonstrate that a remarkably good approximation to the
 the Ter-Mikayelian 
effect can be obtained by approximating
$m_{g}(k)\approx m_{\infty }$ (see also Appendix B1).

\bigskip

\begin{center}
\vspace*{5.cm}
\includegraphics{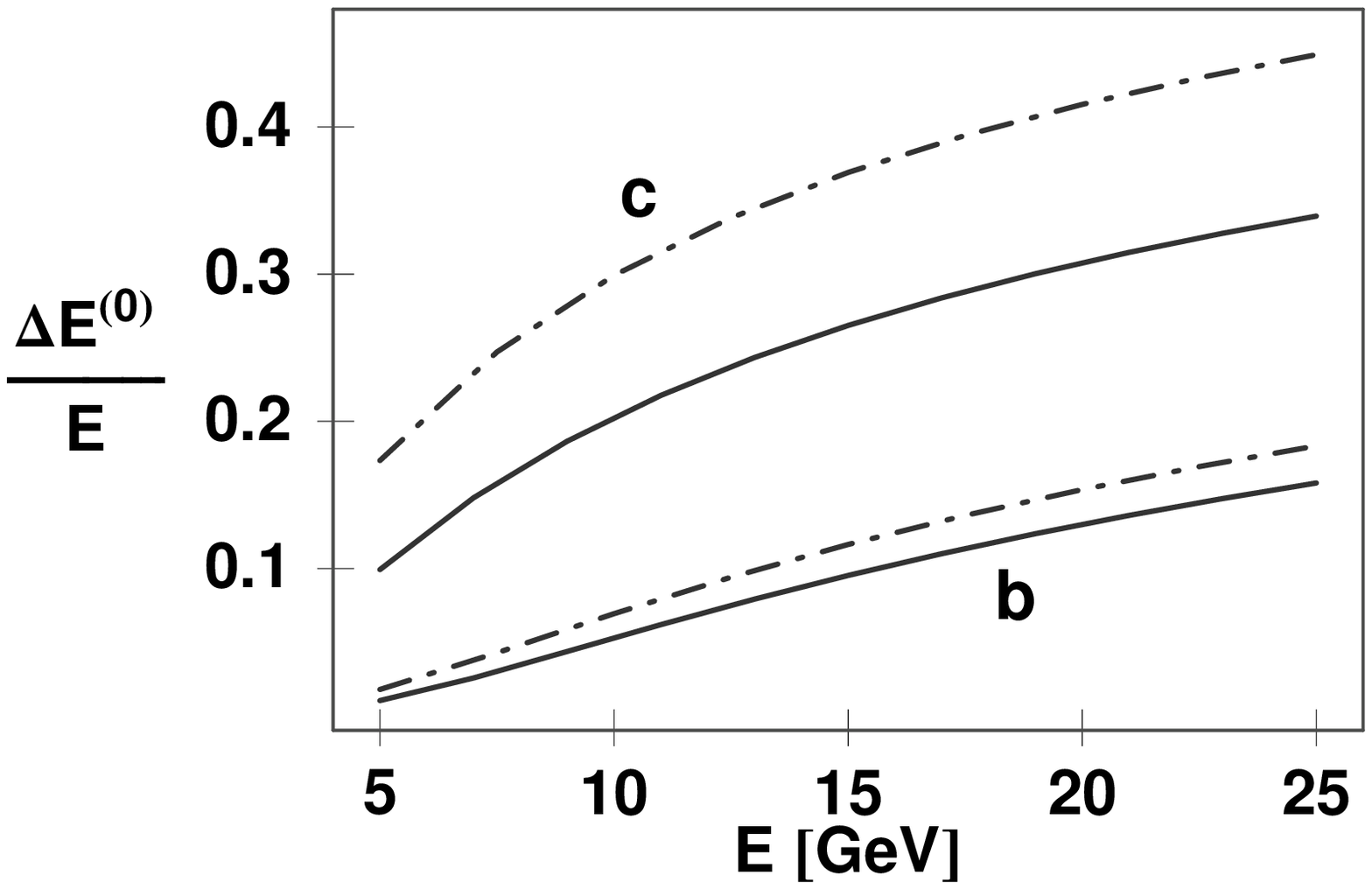}
\vskip 38pt
\begin{minipage}[t]{15.0cm}
{\small {FIG.~9} The reduction of the fractional 
associated vacuum energy loss for charm and bottom 
quark due to the QCD Ter-Mikayelian plasmon effect is shown as a function of 
the quark energy. The upper (lower) dashed-dotted curve shows the vacuum 
energy loss for charm (bottom) quark if gluons are treated as massless and 
transversely polarized.  The upper (lower) solid curve shows the effect of using
the medium dispersion for gluons.  We take $\mu=0$ GeV for the vacuum 
and $\mu=0.5$ GeV for the medium cases.}
\end{minipage}
\end{center}
In Fig. 9 we compare the associated energy loss in the vacuum and a medium with
$\mu=0.5 $ GeV.
We see that for charm quark medium energy loss is significantly reduced by 
 $ \approx 30 \% $. The Ter-Mikayelian plasmon  
effect therefore {\em enhances} the yield of high transverse charm quarks relative 
to the vacuum case. On the other hand, we see that for bottom quark Ter-Mikayelian effect 
is negligible as is the absolute associated radiated energy
because the dead cone~\cite{Dokshitzer:2001zm} is so wide in that case.
 
\begin{center}
\vspace*{6cm}
\includegraphics{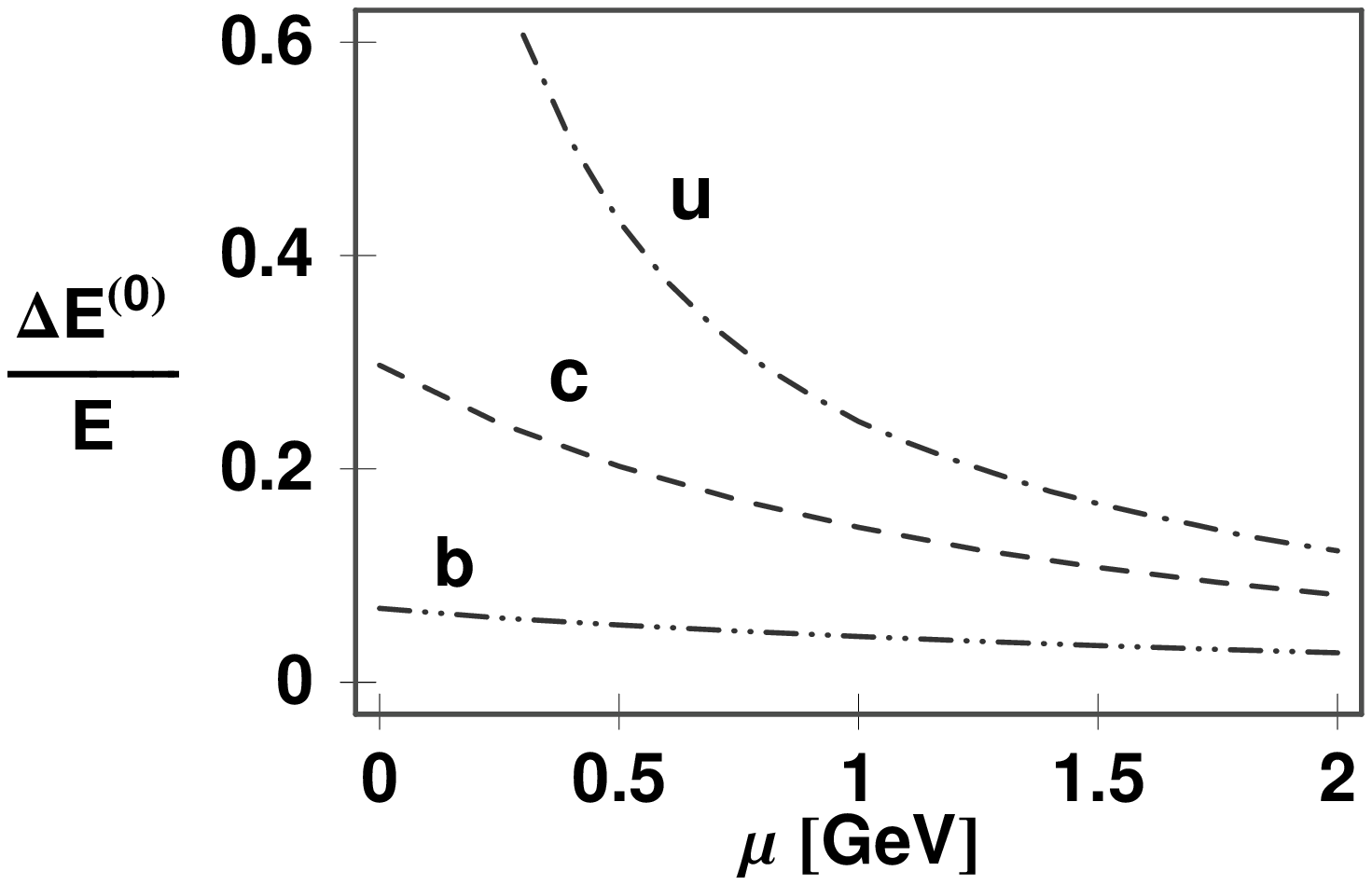}
\vskip 38pt
\begin{minipage}[t]{15.0cm}
{\small {FIG.~10} Medium modified zeroth order in opacity fractional energy 
loss is shown as a function of Debye mass $\mu$ for different mass quarks. 
We see that medium 
effect is more important for light (dot-dashed curve) and charm (dashed curve) 
quark energy loss, since it leads to the strong suppression of the associated 
radiation. On the other hand, bottom quark (dot-dot-dashed curve) energy loss 
shows only a weak dependence on the medium.}
\end{minipage}
\end{center}

So far, we have concentrated on the Ter-Mikayelian effect for only heavy 
quarks. The generalization of the plasmon effect to light quarks is not 
trivial due to the fact that the light quark vacuum energy loss is infrared 
divergent. However, as shown in Fig.10 we see that in a QCD medium, dynamic 
polarization naturally regulates infrared divergences for light quarks, since 
both quarks~\cite{Kapusta,Le_Bellac} and gluons acquire a finite self 
energy. Confinement in the vacuum naturally limits the effective screening 
scale to  $\mu_{vac}> \Lambda_{QCD}$. 

\section{Conclusion}

In this paper we computed both the transverse and longitudinal contributions 
to the lowest order in opacity radiative quark energy loss. We have 
shown that longitudinal contribution can be neglected for the energy range of 
experimental interest. We have also seen that transverse polarization can be, 
for moderate range of temperatures ($0.5 \le \mu \le 1$ GeV), 
approximated by simple form $D_{\mu \nu }
\approx -\frac{P_{\mu \nu }}{k^{2}-m_{g}^{2}}$, where $P_{\mu \nu }$ is 
transverse projector and $m_{g}\approx m_\infty=\mu/ \sqrt{2}$. It is 
remarkable how well the effect of medium polarization $\Pi_{\mu, \nu} 
(\omega, \vec {\bf k})$ can be approximated numerically in this simple way.
In a subsequent paper~\cite{Gyulassy_Djordjevic}, 
we will use this approximation to simplify the calculations of 
medium induced radiative energy loss.
\medskip

In the appendix B it was also seen that the poles of the gluon propagator 
give both the contribution from quasi-particles and "particle hole 
excitations" (which correspond to the energy $\omega$ smaller than 
momentum $|\vec {\bf k}|$). However, we have shown that the particle hole 
contribution to the radiative energy loss is negligible compared to the 
radiative one.

\medskip

The next step beyond the 1 loop HTL approximation employed here will
be to compute the 2 loop (1st order in opacity corrections) to the
associated energy loss. Cutting two HTL loops diagrams, we pick up
three different contributions to the energy loss. Two of them give
corrections to the plasmon mass and particle hole energy loss
respectively. These are perturbative higher order corrections to the
results here. However, a third contribution, in which the poles of two HTL
propagators give one quasi-particle and one particle-hole excitation give rise 
to additional energy loss. This contribution corresponds to the 
generalization of the first order opacity correction to medium induced 
radiative energy loss computed for light partons in~\cite{GLV}. The results 
were summarized in I~\cite{DG-PLBCharm}, and the details of that computation 
will be reported in a subsequent paper~\cite{Gyulassy_Djordjevic}.

\vspace{1.0 cm}

{\em Acknowledgments:} Valuable discussions with I. Vitev, Z. Lin, J.
Nagle, and W. Zajc on heavy quark production at RHIC are gratefully
acknowledged. This work is supported by the Director, Office of
Science, Office of High Energy and Nuclear Physics, Division of
Nuclear Physics, of the U.S. Department of Energy under Grant No.
DE-FG02-93ER40764.

\begin{appendix}

\section{Gluon propagator in temporal axial gauge}

We recall here some of the basic tensorial properties of the gluon 
propagator, in axial gauges ($u^{\mu} A_{\mu}(\vec{\mathbf{x}},t)=0$, 
$u^\mu$ a fixed four vector). 

The transverse projector $P^{\mu\nu}$ (with respect to both $k^\mu$ 
and $u^\mu$), is given by Eq. (\ref{Pmunu}). Note that $\overline{u}^{2}=u^2-(ku)^2/k^2$ is well defined even in light cone $(u^2=0)$ gauge.

The orthogonal longitudinal projector, $Q^{\mu\nu}$ (with respect to $k^\mu$),
is given by
 
\beqar
Q_{\mu \nu }&=&(g_{\mu \nu }-\frac{u_{\mu }k_{\nu }+
k_{\mu }u_{\nu }}{(u\cdot k)}+\frac{u^{2}k_{\mu }k_{\nu }}{(u\cdot k)^{2}}-
P^{\mu\nu})\frac{(u \cdot k)^{2}}{k^{2}u^{2}}
\nonumber \\
&=& \left( \frac{u^\mu u^\nu}{u^2} (uk)^2 + k^\mu k^\nu u^2 - (u^\mu k^\nu+k^\mu u^\nu)(uk)\right) \frac{1}{u^2 k^2 -(uk)^2} 
\eeqar{qmn}

In terms of these projectors the free gluon propagator in the axial gauge is

\beqar
D^{(0)\mu\nu}(k) &=&-\Delta(k)\left( P^{\mu\nu} + \beta(k,u)  
Q^{\mu\nu} \right) \nonumber \\
&=& -\frac{1}{k^2+i\epsilon} \left( g^{\mu\nu}- \frac{u^\mu k^\nu +
 k^\mu u^\nu}{(uk)}+ u^2 \frac{k^{\mu} k^\nu}{(uk)^2} \right)
\eeqar{d0mn}
where 
\beq
\beta(k,u) =\frac{k^2 u^2}{(uk)^2} 
\;\;. \eeq{beta} 

is a kinematic factor. In the temporal axial gauge, $u=(1,0,0,0)$, $\beta=1-\vk^2/\omega^2$ and the projectors reduce to 

\beqar
&&P_{\mu \nu }=Q_{\mu \nu }=0, \; {\rm if} \; \mu , \; \nu =0 \nonumber \\
&&P_{ij}=-\delta_{ij} + \frac{k_{i}k_{j}}{\overrightarrow{k}^{2}}, \;
Q_{ij}=-\frac{k_{i}k_{j}}{ \overrightarrow{k}^{2}}, \; {\rm if} \; i,j=1,2,3.
\eeqar{} 

In a medium with four velocity $u^\mu$ the gluon acquires a
temperature dependent self energy $\Pi_{\mu \nu }$ in addition
to its vacuum self energy. The one-loop (Hard Thermal Loop~\cite{Rebhan} (HTL) or equivalently, eikonal 
linear response~\cite{Gyulassy_Selikhov}) self energy can be decomposed as

\beq    
\Pi_{\mu \nu }=\Pi_{L}R_{\mu \nu }+\Pi_{T}P_{\mu \nu }
\eeq{}

where 

\beq
R_{\mu \nu } =
\frac{\overline{u}_{\mu } \overline{u}_{\nu }}{\overline{u}^{2}}
\eeq{rdef} 

is the longitudinal projector in covariant gauges with respect to $u^\mu$ 
but transverse with respect to $k^\mu$. Therefore, the HTL self
energy is transverse with respect to $k$, $k\Pi=\Pi k=0$.
Note that $RP=PR=0$, $R^2=R$, and 
\beq
QRQ=Q/\beta(k,u)
\;\;. \eeq{qrq} 
where $\beta$ is given by (\ref{beta}).

The HTL medium modified gluon propagator ($D_{\mu \nu }$) can be obtained
 by solving the Dyson equation~\cite{Kapusta}
\beqar
D_{\mu \nu }&=& D_L Q_{\mu\nu}+ D_T P_{\mu\nu}= D^{(0)}_{ \mu \nu }-
D^{(0)}_{ \mu \alpha }
\Pi^{\alpha \beta }D_{\beta \nu }
\nonumber \\
&=& -\frac{P_{\mu \nu }}{\omega^{2} \epsilon_{T} - 
\vec{\mathbf{k}}^{2}} - \frac{Q_{\mu \nu }}{\omega^{2} \epsilon_{L}}
\;\; .
\eeqar{dyson}

Note that, in general~\cite{Gyulassy_Selikhov}, we would have one extra 
term $\eta \frac{k_{\mu}k_{\nu}}{k^{4}}$ in the gluon propagator, where $\eta$ 
is a gauge parameter. However, in temporal axial gauge~\cite{Kapusta},
 
$D_{\mu \nu}=0$, if $\mu$, $\nu=0$.

\bigskip

Therefore, in temporal axial gauge $\eta$ has to be equal to zero, i.e. 
extra term vanishes.

\section{Cutkosky rules for gluon propagator in the medium}

In this appendix we want to justify using the Cutkosky rules for gluon 
propagator in the medium, which is represented by Eq.~(\ref{Dmunu}) in our 
computations.

\bigskip

Diagram $M$

\begin{center}
\vspace*{2.3cm}
\includegraphics{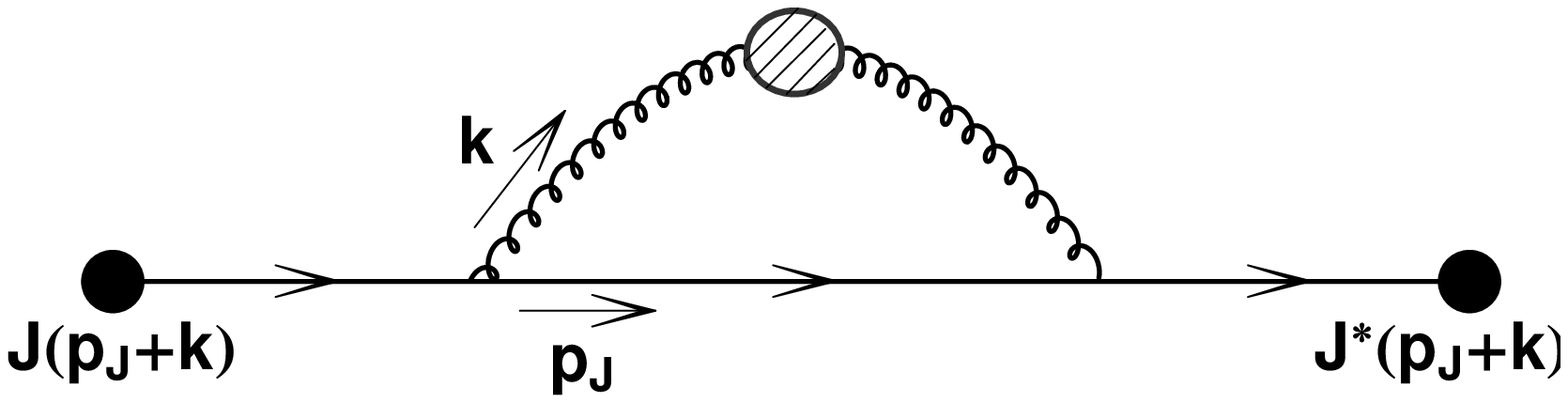} 
\hspace{-1.7cm}
FIG.~11 
\end{center}

corresponds to $\sum M_{n}$, where $M_{n}$ is the amplitude of the following 
diagram:

\begin{center}
\vspace*{5.8cm}
\includegraphics{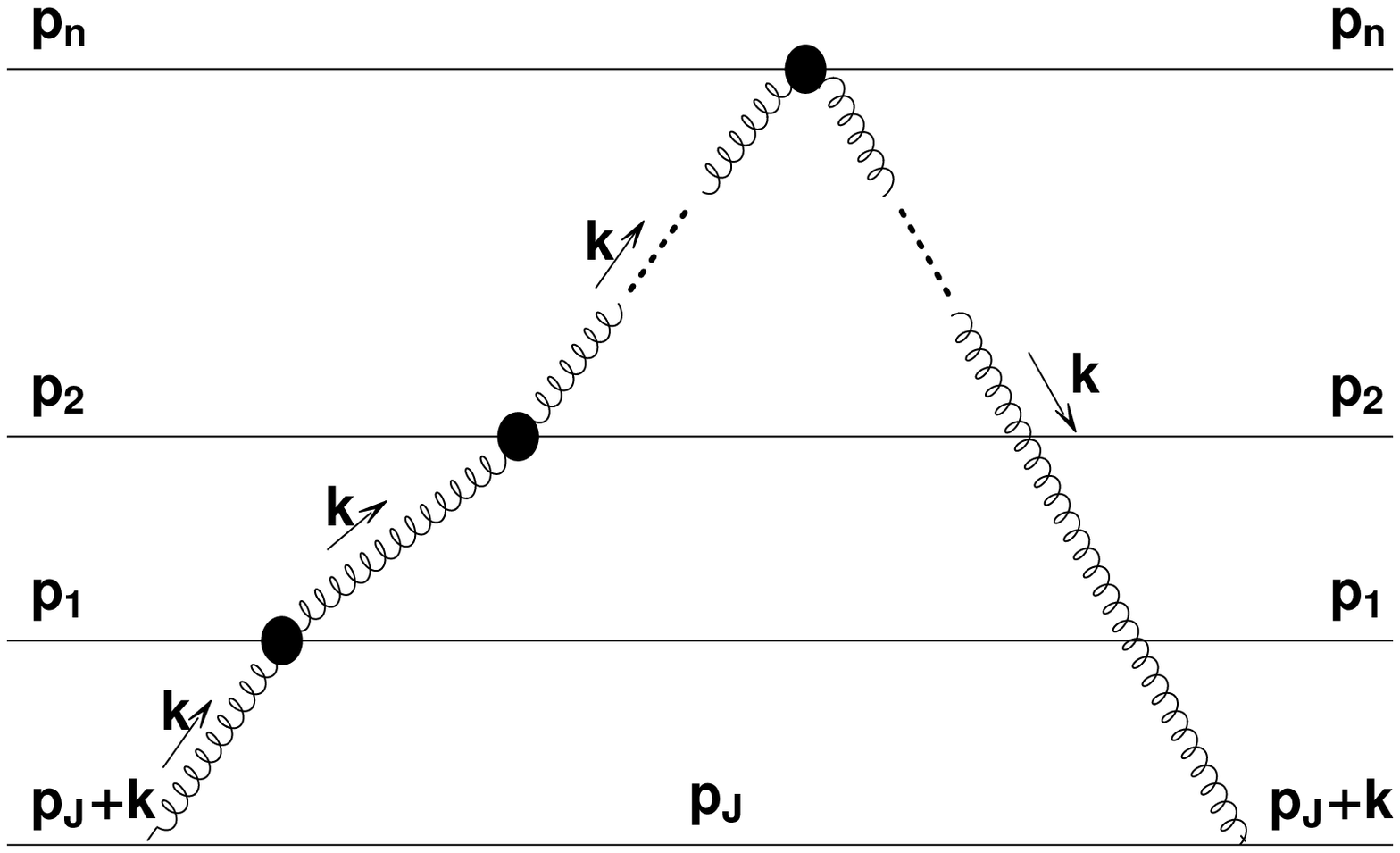}  
\vskip 25pt
\hspace{-1.0cm}
FIG.~12
\end{center}
\vskip 15truemm

Here,

\begin{center}
\vspace*{3.0cm}
\includegraphics{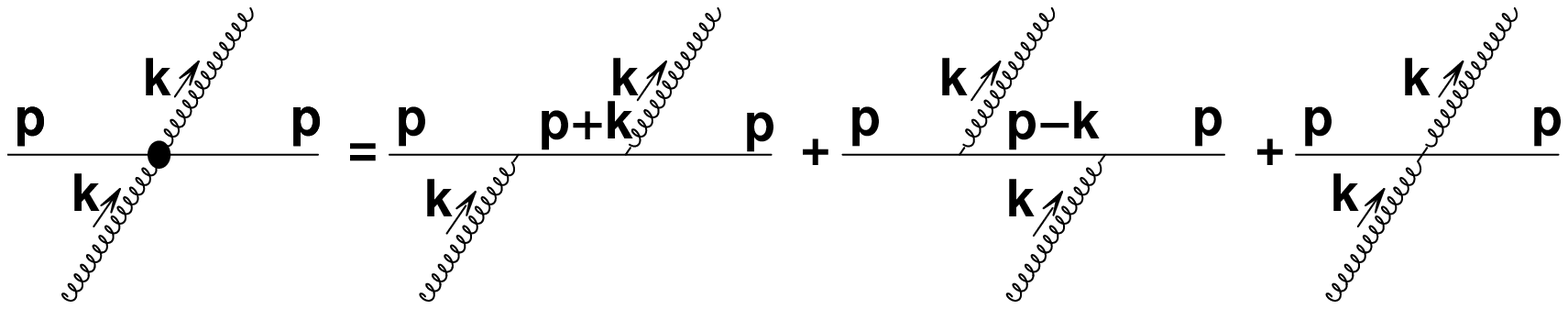}  
\vskip 0pt
FIG.~13
\end{center}

\beqar
&\; &\langle iM_{n}\rangle =Tr \{ \sum_{a_{0}} \int \frac{d^{4}p_{J}}
{(2\pi )^{4}}
\frac{d^{4}k}{(2\pi )^{4}}|J(p+k)|^{2}\frac{1}
{((p_{J}+k)^{2}-M^{2}+i\epsilon )^{2}}
\frac{i}{p_{J}^{2}-M^{2}+i\epsilon } \nonumber \\[1ex]
 && \times (-ig_{s}(2p_{J}+k)^{\mu _{0}}T_{a_{0}})
\frac{-ig_{\mu_{0}\nu_{1}}}{k^{2}+i\epsilon } i\pi^{\nu_{1}\mu_{1}}
\frac{-ig_{\mu _{1}\nu_{2}}}{k^{2}+i\epsilon} i\pi ^{\nu _{2}\mu _{2}}...
i\pi ^{\nu _{n}\mu _{n}} \frac{-ig_{\mu _{n}\nu _{0}}}{k^{2}+i\epsilon }
(-ig_{s}(2p_{J}+k)^{\nu_{0}}T_{a_{0}})\}
\eeqar{M_n}

where $\pi^{\mu \nu } (k)$ is the amplitude of the diagram shown on Fig.13, 
averaged over thermal momentum distribution $n_{eq}(p)$ given by Eq. 
(\ref{n(p)}).

\beqar
i\pi^{\mu \nu } (k)&=& \int d^{4}p n_{eq}(p)\{\frac{i}{(p+k)^{2}+i\epsilon }
(-ig_{s}(2p+k)^{\mu })(-ig_{s}(2p+k)^{\nu })+  \nonumber \\[1ex]
 &\;& + \frac{i}{(p-k)^{2}+i\epsilon }(-ig_{s}(2p-k)^{\mu})
(-ig_{s}(2p-k)^{\nu })+2ig^{\mu \nu }\} \sum_{j} jT_{a}T_{b} j^{\dagger }
\eeqar{pi}

where $ \sum_{j} jT_{a}T_{b}j^{\dagger } = \frac{1}{2}\delta_{ab}$, 
and $n_{eq}(p)$ is the equilibrium momentum distribution ~\cite{Gyulassy_Selikhov} at temperature $T$ including both quarks and gluons 

\beq
n_{eq}(p)=\frac{1}{2}(Q_{eq}^{+}+Q_{eq}^{-})+NG_{eq}.
\eeq{n(p)}

Here

\beq
Q_{eq}^{\pm} = \frac{2N_{f}}{(2\pi)^{3}} 2\theta(\pm p^{0})
\delta (p^{2}) (\exp (\pm \frac{p^{0}}{T})+1)^{-1}
\eeq{}

and

\beq
G_{eq} = \frac{2N}{(2\pi)^{3}} 2 \theta(p^{0}) \delta (p^{2})
(\exp (\frac{p^{0}}{T})-1)^{-1}
\eeq{}

are quark (antiquark) and gluon distributions respectively. $N_{f}$ is the 
number of flavors, and $N$ is the number of colors.
 
In the soft gluon limit $\pi^{\mu \nu}$ becomes
 
\beq
\pi^{\mu \nu }= -g_{s}^{2} \int d^{4}p \frac{p^{\mu }k_{\alpha }(p^{\alpha}
\partial_{p}^{\nu} - p^{\nu } \partial_{p}^{\alpha })} {(pk)+i\epsilon }
n_{eq}(p),
\eeq{PmunuSoft}

in agreement with~\cite{Gyulassy_Selikhov}.

\medskip

It is easy to prove that $\pi^{\mu \nu }$ is transverse, i.e. 
$\pi^{\mu \nu } k_{\nu } = k_{\mu } \pi^{\mu \nu }=0 $.

\bigskip 

Therefore, in covariant gauge $\pi^{\mu \nu }$ can be decomposed:

\bigskip 

\beq
\pi^{\mu \nu } = \Pi_{T} P^{\mu \nu } + \Pi_{L} R^{\mu \nu},
\eeq{dec1}

where $P^{\mu \nu }$ and $R^{\mu \nu }$ are given by Eqs. (\ref{Pmunu}) and 
(\ref{rdef}) respectively.

\medskip

Using Eqs.~(\ref{PmunuSoft}, \ref{dec1}, \ref{Pmunu}, \ref{rdef}) we get

\beq
\Pi_{L} = Q_{\mu \nu } \pi^{\mu \nu } = -g_{s}^{2} 
\frac{k^{2}}{u^{2}k^{2}-(ku)^{2}} \int d^{4}p (pu) 
\frac{(pk)(u\partial n)-(pu)(k\partial n) }{(pk)}
\eeq{Pi_L}

\beq
\Pi_{T} = \frac{1}{2} P_{\mu \nu } \pi^{\mu \nu } = -\frac{1}{2}
g_{s}^{2} \int d^{4}p \frac{1}{(pk)} \{(k\partial n)(\frac{k^{2}(pu)^{2}}
{u^{2}k^{2}-(ku)^{2}}-p^{2})+(pk)((p\partial n)-\frac{k^{2}(pu)(u\partial n)}
{u^{2}k^{2}-(ku)^{2}}) \}
\eeq{Pi_T}

By replacing $n_{eq}(p)$ from Eq.~(\ref{n(p)}) we get 

\bigskip

\beq
\Pi_{L}= - \frac{(\omega ^{2} - \vec{\mathbf{k}}^{2}) \mu^{2}}
{\vec{\mathbf{k}}^{2}}
(1+\frac{\omega }{2|\vec{\mathbf{k}}|} 
\log |\frac{\omega -|\vec{\mathbf{k}}|}{\omega +|\vec{\mathbf{k}}|}|)
\eeq{}

and

\beq
\Pi_{T} = \frac{\mu^{2}}{2} + \frac{(\omega ^{2} - \vec{\mathbf{k}}^{2})
\mu^{2}}{2 \vec{\mathbf{k}}^{2}} (1+\frac{\omega }{2|\vec{\mathbf{k}}|}
\log |\frac{\omega -|\vec{\mathbf{k}}|}{\omega +|\vec{\mathbf{k}}|}|)
\eeq{}
\bigskip 

where $\mu=g_{s}T\sqrt{1+\frac{N_{f}}{6}}$.

\bigskip

These results are in agreement with~\cite{Gyulassy_Selikhov}, and they lead 
to $ \epsilon_{L}$ and $\epsilon_{T}$ which we use in this paper.

\bigskip 

Using $\pi^{\mu \nu }$ from Eq.~(\ref{dec1}) $M$ finally becomes:

\beqar
M &=& (-i) \int  D_{R} |J(p_{J})|^{2} \frac{d^{4}p_{J}}{(2\pi )^{4}} 
\frac{1}{p_{J}^{2}-M^{2}+i\epsilon } \times \nonumber \\[1ex]
 &\;& \times \int C_{R} g_{s}^{2} \frac{d^{4}k}{(2\pi )^{4}} \frac{1}{((p_{J}+k)^{2}-M^{2}+i\epsilon )^{2}}
(2p_{J}+k)^{\mu }D_{\mu \nu}(2p_{J}+k)^{\nu}
\eeqar{amplitudeM}

where $D_{\mu \nu } = -\frac{P^{\mu \nu }}{k^{2}-\Pi _{T}+i \epsilon} - 
\frac{Q^{\mu \nu }}{k^{2}-\Pi_{L}+i \epsilon}$, and we assume that $\epsilon$ 
is positive.

\bigskip 

Lets now compute

\beq
\int d\omega \frac{1}{((p_{J}+k)^{2}-M^{2}+i\epsilon )^{2}}
(2p_{J}+k)^{\mu} D_{\mu \nu }(2p_{J}+k)^{\nu }=\int d\omega (\frac{f_{T}(p_{J},k)}{k^{2}-\Pi _{T}+i \epsilon} - 
\frac{f_{L}(p_{J},k)}{k^{2}-\Pi_{L}+i \epsilon})]
\eeq{int_omega}

Since we are interested only in the radiative energy loss, we assume that 
initial jet is off-shell, and therefore we have 
that $f_{T}(p_{J},k)$ and $f_{L}(p_{J},k)$ are analytic functions.
 
\bigskip 

>From Eqs.~(\ref{Pi_L}, \ref{Pi_T}) we see that $\Pi_{T}$ and $\Pi _{L}$ can be
written as $ \sum_{p} \frac{ \xi_{T(L)}(p,k)}{(pk)}$ where $\xi_{T(L)}(p,k)$ 
are analytic functions, and where we have assumed that we have
discrete set of $p's$.

\bigskip 

Since discussion for both parts of the integral (\ref{int_omega}) is the 
same, we will consider only one part:

\beq
\int d\omega \frac{f(p_{J},k)} {\omega^{2}-\vec{\mathbf{k}}^{2} - 
\sum_{p} \frac{\xi (p,k)}{(pk)}+i \epsilon}.
\eeq{contour_int}

Note that $\frac{\xi (p,k)}{(pk)}$ is infinite when 
$\omega =\frac{\vec{\mathbf{p}}\cdot \vec{\mathbf{k}}}{E_{\mathbf{p}}}$.

\bigskip

For simplicity, suppose that all 
$\frac{\vec{\mathbf{p}}\cdot\vec{\mathbf{k}}}{E_{\mathbf{p}}}$ are different. We can order them in such a way that 
$\omega _{1}<\omega _{2}<...<\omega _{n}\leq |\vec{\mathbf{k}}|$.

Than, for fixed $|\vec{\mathbf{k}}|$ zeros, $\pm(\Omega_{i}-i\epsilon)$, 
of $(\omega^{2} -\vec{\mathbf{k}}^{2}-\sum_{p} \frac{\xi(p,k)}{(pk)}+
i\epsilon)$ can be found graphically:
 
\begin{center}
\vspace*{8.0cm}
\includegraphics{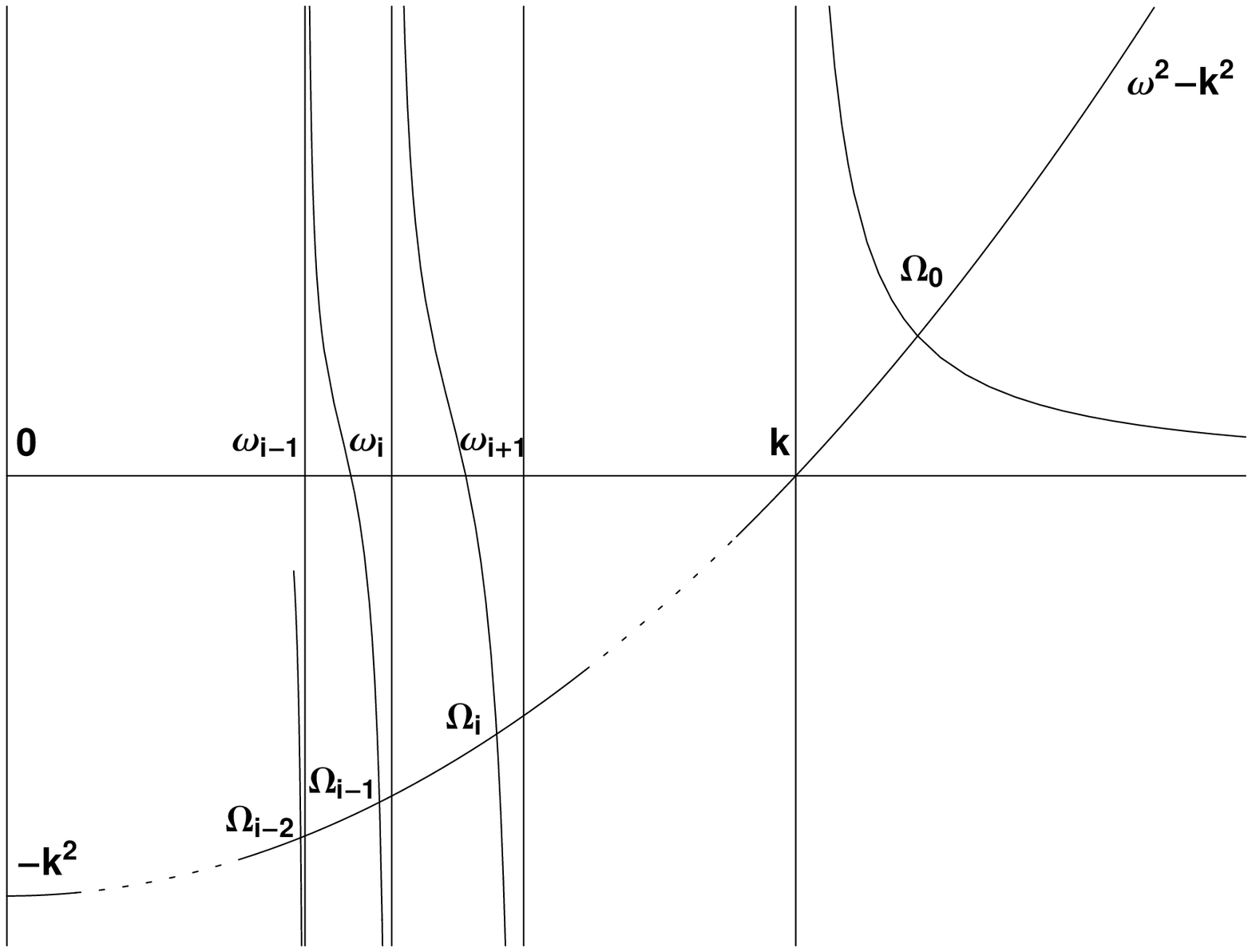}  
\vskip -10pt
\begin{minipage}[t]{15.0cm}
{\small {FIG.~14} shows schematically the zeros $\Omega_{i}$ for $\omega^{2} 
- \vec{\mathbf{k}}^{2}-\sum_{p} \frac{\xi (p,k)}{(pk)}$ at fixed $|\vec{\mathbf{k}}|$.} \end{minipage}
\end{center}

We see that we have $n^{\prime}<n$ different solutions $\Omega_{i} \leq 
|\vec{\mathbf{k}}|$, and exactly one solution $\Omega_{0} > 
|\vec{\mathbf{k}}|$. Then, if we close the contour in Eq. (\ref{contour_int}) 
in clockwise direction, we will pick up only positive poles, i.e. 
$(\Omega_{i}-i\epsilon)$.   

\bigskip 

Therefore,

\beq
\int d\omega \frac{f(p_{J},k)}{\omega^{2} - \vec{\mathbf{k}}^{2} - 
\sum_{p} \frac{\xi (p,k)}{(pk)}+ i \epsilon} =(- 2 \pi i) 
\sum_{i=1}^{n^{\prime}} Res(\Omega_{i}) +(- 2 \pi i) Res(\Omega_{0}).
\eeq{}

In the following subsection, we will test the relative magnitude of the two 
contributions for the case of the dominant transverse excitations.
Solutions $\Omega_{i}<$\bigskip $|\vec{\mathbf{k}}|$ correspond to particle 
hole excitation, and we will prove that this contribution 
is negligible.

\subsection{Simplifying the gluon propagator in hot dense medium}

The transverse response 

\beq
\rho_{T}(k) \equiv \frac{1}{2 \pi} Disc [\frac{1}{k^2-\Pi_{T}(k)} ]
\eeq{}
 
obeys the sum rule~\cite{Le_Bellac}

\beq 
\int_{-\infty}^{\infty} d \omega \omega \rho_{T}(\omega,|\vec{\mathbf{k}}|)=1
\eeq{sum_rule}

\bigskip

We can write

\beq
\rho_{T}(\omega, |\vec{\mathbf{k}}|)=\beta(\omega, |\vec{\mathbf{k}}|)+
\delta(\omega^{2}\epsilon_{T}-\vec{\mathbf{k}}^{2}),
\eeq{ro}

where first part in this formula corresponds to the particle hole excitation, 
and the second part corresponds to the delta function contribution.

It is easy to show that

\beq
\delta(\omega^{2}\epsilon_{T}-\vec{k}^{2})= 
\delta (\omega^{2} -\omega _{T}^{2}) Z_{T}(k)
\eeq{}

where $\omega _{T}$ is the transverse plasmon spectrum, and 

\beq
Z_{T}(k)=\frac{2 \omega_{T} ^{2}(\omega_{T} ^{2}-\vec{\mathbf{k}}^{2})}
{\omega_{T} ^{2}\mu^{2} - (\omega_{T} ^{2}-\vec{\mathbf{k}}^{2})^{2}}.
\eeq{}

The sum rule reduces to

\beq
Z_{T}(k)+\int_{-\infty}^{\infty} d \omega \omega \beta(\omega,|\vec{\mathbf{k}}|)=1
\eeq{sum_rule_final}

\begin{center}
\vspace*{5.5cm}
\includegraphics{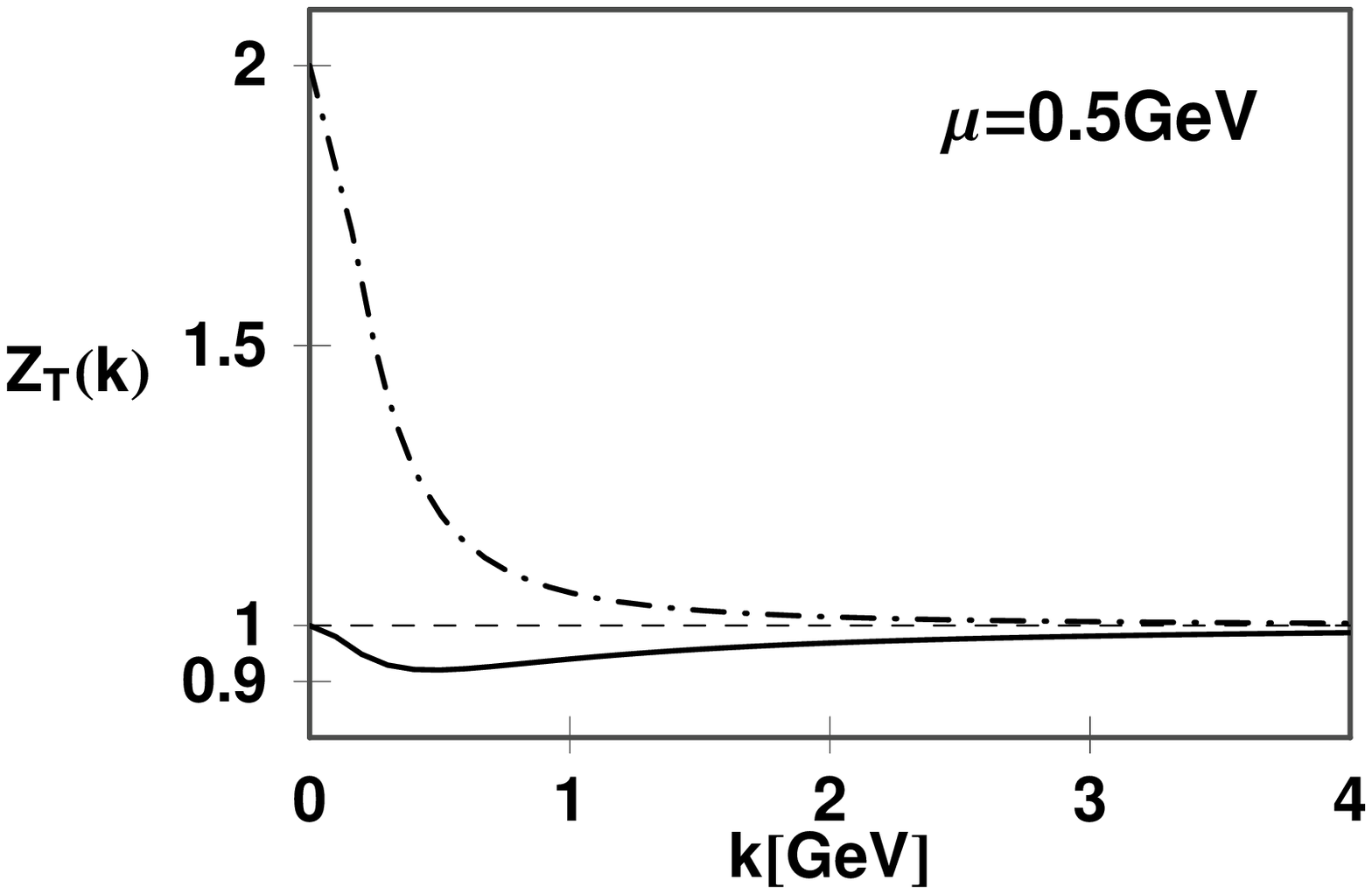} 
\vskip 20pt
\begin{minipage}[t]{15.0cm}
{\small {FIG.~15} shows $Z_{T}(k)$ as a function 
of $|\vec{\mathbf{k}}|$. Full curve shows exact 
$Z_{T}(\omega_{T},|\vec{\mathbf{k}}|)$, while dot-dashed curve shows 
$Z_{T}(\sqrt{\vec{\mathbf{k}}^{2}+m_{g}^{2}},|\vec{\mathbf{k}}|)$, in which 
$\omega_{T}$ is approximated by $\sqrt{\vec{\mathbf{k}}^{2}+m_{g}^{2}}$. }
\end{minipage}
\end{center}
\vskip 4truemm 

In Fig.15 we see that $Z_{T}(\omega_{T},|\vec{\mathbf{k}}|)$ is 
approximately equal to 1 for whole region of $|\vec{\mathbf{k}}|$. Therefore, 
we can conclude that the contribution form the particle hole excitation  is 
negligible.

\bigskip

In order to justify this conclusion we will estimate the particle hole 
contribution to the energy loss. For simplicity, we will consider quarks 
to be massless. 

\medskip

>From Eq. (\ref {yield}) we see that 

\beqar
\frac {\Delta E^{\rm hole}}{E} &=& \int \frac{1}{E} \omega d\omega \; d\cos 
\theta \; \vec{\mathbf{k}}^{2} d|\vec{\mathbf{k}}| \; 
\frac{C_{R} \alpha_{S}}{\pi} \;
\frac{1}{(Q^{2}-M^{2})^{2}} \; 4\vec{\mathbf{p}}^{2}\sin^{2}\theta  \;
\beta(\omega, |\vec{\mathbf{k}}|) \; <
\nonumber \\ 
&<& \frac{2}{E} \frac{C_{R} \alpha_{S}^{\rm max}}{\pi} \int 
\vec{\mathbf{k}}^{2} d|\vec{\mathbf{k}}| \; x dx \; 
\beta(x, |\vec{\mathbf{k}}|)\; ({\rm Ln} (\frac{1+x}{1-x})-2)
\eeqar{hole_contrib}

where $x \equiv \frac{\omega}{|\vec{\mathbf{k}}|}$.

\medskip

\beqar
({\rm Ln} (\frac{1+x}{1-x})-2) \; < \; ({\rm Ln} (\frac{2}{0.05})-2) - 
{\rm Ln}(1-x) \; \Theta(x-0.95)
\eeqar{Ln}

Using Eq. (\ref {Ln}) we get:

\beqar
\frac {\Delta E^{\rm hole}}{E} &<& \frac{2}{E} 
\frac{C_{R} \alpha_{S}^{\rm max}}{\pi} \; \{ \; ({\rm Ln} (\frac{2 }{0.05})-1) 
\int d|\vec{\mathbf{k}}| \; \int \vec{\mathbf{k}}^{2} \;  x dx \;
\beta(x, |\vec{\mathbf{k}}|)\; - \nonumber \\ 
&& \; \; \; \; \; \; \; \; \; \; \; \; \; \; \; \; 
- \int_{0.95}^{1} x  dx \; {\rm Ln} (1-x) \;
\int_{0}^{|\vec{\mathbf{p}}|} d|\vec{\mathbf{k}} | \; \vec{\mathbf{k}}^{2} \;
\beta(x, |\vec{\mathbf{k}}|) \; \}  \; < \nonumber \\ 
&<& \frac{2}{E} \frac{C_{R} \alpha_{S}^{\rm max}}{\pi} \; \{ \; 2 
\int_{0}^{|\vec{\mathbf{p}}|} d|\vec{\mathbf{k}}| \frac{(1-Z_{T}(k))}{2}- \int_{0.95}^{1} x  dx \; {\rm Ln} (1-x)
\int_{0}^{|\vec{\mathbf{p}}|} d|\vec{\mathbf{k}} | \; \vec{\mathbf{k}}^{2} \;
\beta(x, |\vec{\mathbf{k}}|) \; \}
\eeqar{hole_contrib1} 

where in the first integral we have used Eq. (\ref {sum_rule_final}). 
\bigskip
 
Using the fact that $|1-Z_{T}(k)| \sim \frac{m_{g}^{2}}
{2 (\vec{\mathbf{k}}^{2}+{m_{g}^{2}})}$ we get 

\beq
\frac{2}{E} \frac{C_{R} \alpha_{S}^{\rm max}}{\pi} \; 
\int_{0}^{|\vec{\mathbf{p}}|} d|\vec{\mathbf{k}}| (1-Z_{T}(k))
\approx \frac{C_{R} \alpha_{S}^{\rm max}}{\pi} 
\; \frac{m_{g}}{E} \; \frac{\pi}{2} 
\eeq{}

In the $x \rightarrow 1$ region $\beta(x, |\vec{\mathbf{k}}|)$\cite{Le_Bellac} can be approximated by 

\beq
\beta(x, |\vec{\mathbf{k}}|) \approx 
\frac{m_{g}^{2} \; (1-x)}{(2 \vec{\mathbf{k}}^{2}(1-x)+m_{g}^{2})^2}
\eeq{}

Then, the second integral in Eq. (\ref {hole_contrib1}) can be performed 
analytically, leading to: 

\beq
\frac {\Delta E^{\rm hole}}{E} <  C_{R} \alpha_{S}^{\rm max}\; \frac {m_{g}}{E}
\eeq{}

We see that the particle hole contribution to the fractional energy loss is 
on the order $(m_{g}/E)$. Since $m_{g} \ll E$ we can conclude that this contribution is negligible.

\bigskip

Finally, on Fig.7 we saw that $\omega _{T} \approx 
\sqrt{\vec{\mathbf{k}}^{2}+m_{g}^{2}}$. Dot-dashed curve on Fig.15 represents
 $Z_{T}(\sqrt{\vec{\mathbf{k}}^{2}+m_{g}^{2}},|\vec{\mathbf{k}}|)$ as 
a function of $|\vec{\mathbf{k}}|$. Again, with the exception of the small 
$|\vec{\mathbf{k}}|<1$ GeV region,  we see that 
$Z_{T}(\sqrt{\vec{\mathbf{k}}^{2}+m_{g}^{2}},|\vec{\mathbf{k}}|)$ is 
approximately equal to 1 for whole region of $|\vec{\mathbf{k}}|$. Thus, 
we can approximate

\beq
\delta {(\omega ^{2}\epsilon _{T}-\vec{\mathbf{k}}^{2})} \approx
\delta (\omega^{2} -(\vec{\mathbf{k}}^{2}+m_{g}^{2})).
\eeq{}

Since longitudinal contribution is negligible, we can conclude 
that a gluon propagator in a hot dense medium can be approximated by

\beq
D_{\mu \nu } \approx -\frac{P_{\mu \nu }}{k^{2}-m_{g}^{2}+i\epsilon}.
\eeq{}

\end{appendix}

\end{document}